\documentclass[a4paper, 11pt]{article}

\usepackage{amsfonts}
\usepackage{amsmath,amsthm,amssymb,amscd,epsfig}
\usepackage{graphicx}
\usepackage{subfigure}
\usepackage{psfrag}
\newcommand {\N}{\mathbb{N}}
\newcommand {\Z}{\mathbb{Z}}
\newcommand {\R}{\mathbb{R}}

\newcommand {\rd}{\mathrm{d}}
\newcommand {\rO}{\mathrm{O}}
\newcommand {\T}{\mathbb{T}}

\newtheorem{theorem}{Theorem}[section]

\newtheorem{corollary}[theorem]{Corollary}

\numberwithin{equation}{section}

\begin{document}

\title{On (in)elastic non-dissipative Lorentz gases and  the (in)stability of classical pulsed and kicked rotors}
\author{
B. {\ Aguer}\thanks{Benedicte.Aguer@math.univ-lille1.f},
and S. {\ De Bi\`{e}vre}\thanks{Stephan.De-Bievre@math.univ-lille1.fr},
\\
Laboratoire Paul Painlev\'e, CNRS, UMR 8524 et UFR de Math\'ematiques \\
Universit\'e des Sciences et Technologies de Lille \\
F-59655 Villeneuve d'Ascq Cedex, France.\\
Equipe-Projet SIMPAF \\
Centre de Recherche INRIA Futurs \\
Parc Scientifique de la Haute Borne, 40, avenue Halley B.P. 70478 \\
F-59658 Villeneuve d'Ascq cedex, France.
}
\date{\today}
\maketitle
\rightline{\em \small Dedicated to the memory of Pierre Duclos}
\begin{abstract}
We study numerically and theoretically the $d$-dimensional Hamiltonian motion of fast particles through a field of scatterers, modeled by bounded, localized, (time-dependent) potentials, that we refer to as (in)elastic non-dissipative Lorentz gases. We illustrate the wide applicability of a random walk picture previously developed for a field of scatterers with random spatial and/or time-dependence by applying it to four other models.  First, for a periodic array of spherical scatterers in $d\geq2$, with  a smooth (quasi)periodic time-dependence, we show Fermi acceleration: the ensemble averaged kinetic energy $\left<\|p(t)\|^2\right>$ grows as  $t^{2/5}$. Nevertheless, the  mean squared displacement $\left<\|q(t)\|^2\right>\sim t^2$ behaves ballistically. These are the same growth exponents as for random time-dependent scatterers.
Second, we show  that in the soft elastic and periodic Lorentz gas, where the particles' energy is conserved, the motion is diffusive, as in the standard hard Lorentz gas, but with a diffusion constant that grows as $\|p_0\|^{5}$, rather than only as $\|p_0\|$.  Third, we note the above models can also be viewed as  pulsed rotors:  the latter are therefore unstable in dimension $d\geq 2$. Fourth, we consider kicked rotors, and prove them, for sufficiently strong kicks, to be unstable in all dimensions with  $\left<\|p(t)\|^2\right>\sim t$ and $\left<\|q(t)\|^2\right>\sim t^3$.  Finally, we analyze the singular case  $d=1$, where  $\left< \|p(t)\|^2\right>$ remains bounded in time for time-dependent non-random potentials whereas it grows at the same rate as above in the random case.
\end{abstract}

\section{Introduction}
In this paper, we shall be interested in the classical dynamics generated by Hamiltonians of the form
\begin{equation}\label{eq:H}
H(q,p,t)=\frac{p^2}{2}+\lambda V(q,t),
\end{equation}
where the potential $V$ is periodic in its (dimensionless) spatial variable $q$ and periodic or quasi-periodic in its (dimensionless) time variable $t$, with a smooth time-dependence.  
More specifically, we consider potentials of the form
\begin{equation}\label{eq:V}
V(q,t)=\sum_{m\in\Z^d} W(q-x_m,\omega t+\phi_0),\quad x_m=\sum m_i\mathrm{e}_i,
\end{equation}
where the $e_i, i=1\dots d$ form a basis of $\R^d$, with $\| e_i\| = 1$ and where $W(q, \phi)$ is a function on $\R^d\times \T^m$, which is spherically symmetric in $q$, and of compact support contained in the ball of radius $1/2$ centered at $q=0$. It is periodic in $\phi$ so that, with $\omega\in \R^m$ a frequency vector, the potential is periodic or quasi-periodic in time. The model can be thought of as a soft inelastic and periodic Lorentz gas, in which the particle with position $q$ and momentum $p$ scatters off periodically placed scatterers, modeled by the identical and localized time-dependent potentials $W$.  We consider an ensemble of particles with fixed given initial energy $\|p_0\|^2>>\lambda$ starting off in a random initial direction from a position close to the origin.  We will give numerical and analytical evidence for the fact that, for $d\geq 2$, the averaged kinetic energy $\left< \|p(t)\|^2\right>$ and mean squared displacement $\left< \|q(t)\|^2\right>$ 
 of  such an ensemble behave like
\begin{equation}\label{eq:d2results}
\left< \|p(t)\|^2\right> \sim t^{2/5}, \qquad \left< \|q(t)\|^2\right> \sim t^2.
\end{equation}
The motion is in this sense ballistic, and this in spite of the fact that the averaged kinetic energy of the particle increases with time, so that the particle undergoes Fermi acceleration. This is, as we will see, a consequence of the fact that the particle turns while traveling. We wish to stress that the power laws in (\ref{eq:d2results}) are identical to the ones valid for {\em random} time-dependent potentials, studied in \cite{adblafp}. In that paper, the motion of a particle in a random potential of the form
\begin{equation}\label{eq:randomV}
V(q,t)=\sum_m \lambda_m W\left(q-x_m, \omega t+\phi_m\right)
\end{equation}
was considered. Contrary to the situation described in (\ref{eq:V}), the scattering centers $x_m$ in (\ref{eq:randomV}), as well as the initial phases $\phi_m$ and the coupling constants $\lambda_m$ are {\rm i.i.d.} random variables. It was shown in \cite{adblafp} that the statistical properties of the motion in such a potential are well described by a random walk; this allows for the prediction of the asymptotic behaviour of the particles' mean position and momentum. In the numerics performed in \cite{adblafp} it was already briefly noted that even if the coupling constants and scattering centers are non-random, so that the only randomness remaining is in the phases $\phi_m$, this asymptotic behaviour is unaltered. We go a step further in the present paper: if the spatial dimension of the system is greater than one, and if the random phases $\phi_m$ are all chosen equal, leading to a purely deterministic potential, periodic in space and (quasi-)periodic in time (as in (\ref{eq:V})), the asymptotics of the particles' motion is not altered. This is, as we will see, a result of the fact that the geometry of the periodic lattice together with the instabilities in the dynamics of the individual scattering events undergone by the particle suffice to effectively randomize the motion, even if the potential itself is completely deterministic: the motion is in this sense self-randomizing.

We will support the results in (\ref{eq:d2results}) with numerical data for two simple models of $W$ on the one hand (Figure~\ref{fig:d2pulsedp}), and with a straightforward extension of the ideas developed in \cite{adblafp} which, as pointed out already, dealt with {\em random} time-dependent potentials. The central idea of this analysis is the claim that the particle motion is efficiently described by a random walk, in which one step corresponds to one crossing by the particle of an effective correlation length of the potential, which is present, as we shall see, in spite of the fact that the potential is not random.  We will show the growth exponents in (\ref{eq:d2results}) are directly related to the typical momentum transfer $\Delta p$ undergone by the particle during one such crossing. In the present situation, it obeys $\Delta p\sim \left\|p\right\|^{-2},$ as we will show.

In order to illustrate the versatility of our approach we mention it also applies to other types of inelastic Lorentz gases that have been considered in the literature. In \cite{lra1,lra2}, the scatterers are hard obstacles, as in the usual elastic Lorentz gas, but with periodically or stochastically moving boundaries.  It is found that the particles' kinetic energy grows linearly in time for the stochastic case. This can also be inferred from our analysis, since, as the authors show, in these systems, the typical momentum transfer in one collision is of order $\|p\|^0\sim 1$, and does therefore not decrease with momentum. As such, these systems are similar to kicked rotors, discussed below, for which the energy growth is indeed linear, as we shall establish. 
For periodically moving boundaries, the authors report a slightly faster energy growth in the systems they study, presumably due to the persistence of correlations; no quantitative analysis is made of the numerically determined exponent, however.  Let us finally mention that no analysis of the behaviour of the particle position is performed in those works, either. We conjecture it to be of the same type as in the kicked rotor, namely $\langle \|q(t)\|^2\rangle\sim t^3$ (see below).

Our random walk analysis  furthermore applies to the soft {\em elastic} Lorentz gas ($d\geq 2$), in which the potentials $W$ in (\ref{eq:V}) are independent of time. Energy is then conserved and hence the kinetic energy of the particle is bounded, and constant in time between collisions. We show that the particle then diffuses in space, so that
\begin{equation}\label{eq:softlorentzelastic}
\langle \|q(t)\|^2\rangle \sim t. 
\end{equation}
In addition, we show the diffusion constant to grow very fast, as $\|p_0\|^5$ and we illustrate these claims with numerical data (Figure~\ref{fig:softlorentz}). Note that in the usual Lorentz gas, with hard scatterers, the motion has been rigorously proved to be diffusive as well~\cite{bs2}, but the diffusion constant then scales as $\| p_0\|$, due to the trivial scaling of the motion with the speed.  We are not aware of rigorous results on the motion through soft elastic scatterers. 

There is a second interpretation of the results described above, to which we now turn. 
There has been continued interest in the study of time-periodic or time-quasiperiodic perturbations of completely integrable Hamiltonian systems, both in classical and quantum mechanics. The basic problem is the one of their stability. Noting that the Hamiltonian in (\ref{eq:H}) is invariant under the lattice translation group 
$\Z e_1+\dots+\Z e_d$ one concludes it generates a Hamiltonian dynamics on $\T^d\times \R^d$, where $\T^d$ designates the torus $\T^d=\R^d/(\Z e_1+\cdots+\Z e_d)$. With this change of viewpoint the Hamiltonian (\ref{eq:H}) therefore describes a particle moving on a flat torus, perturbed by a scatterer modeled by the time (quasi-)periodic potential $W$. Such systems are referred to as pulsed rotors, and are perturbations of the completely integrable system consisting of a free particle on a flat torus.  
They are said to be stable if their kinetic energy is uniformly bounded in time, i.e. $\sup_{t}\|p(t)\|<+\infty$. 

Viewing the Hamiltonian~\eqref{eq:H} as describing a pulsed rotor, the first relation of (\ref{eq:d2results}) shows that pulsed rotors are unstable when $d\geq 2$.
As will become clearer below, the potential $V$ constitutes an increasingly small perturbation of the free particle, as the particle momentum gets larger. The unbounded growth of the energy observed in (\ref{eq:d2results}) therefore suggests that either all invariant tori of the unperturbed system are destroyed by the perturbation, or that the momentum variable diffuses between the remaining tori, slowly increasing with time at a rate given in (\ref{eq:d2results}). We have not explored any further the complicated mechanisms, possibly related to Arnold diffusion, by which this may happen; our results are obtained through a  much simpler analysis of the momentum transfer in a single scattering event, which is sufficient to predict the rate of growth of the diffusion mechanism.

The preceding analysis does not apply when $d=1$, where the motion in a random time-dependent potential as in (\ref{eq:randomV}) is completely different from the one in a potential that is (quasi-)periodic in time as in (\ref{eq:V}). Indeed, we find for the models we study numerically that, for very long times, (see Figure~\ref{fig:pulsedd1})
\begin{equation}\label{eq:d1results}
\left< p^2(t)\right> \sim 1,\quad \left< q^2(t)\right> \sim t.
\end{equation}
This is in contrast with the behaviour in a random potential of the type (\ref{eq:randomV}), where one has (see \cite{adblafp}) for $d=1$: 
\begin{equation}\label{eq:d1randomresults}
\left< p^2(t)\right> \sim t^{\frac25},\quad \left< q^2(t)\right> \sim t^{\frac{12}{5}}.
\end{equation}
We will explain this difference in Section~\ref{s:onedrotor}. Note that (\ref{eq:d1results}) indicates the system to be stable. We will provide two theorems (Theorems~\ref{th:averaging} and \ref{co:approxp}) supporting this last claim for arbitrary long times using basic time-dependent Hamiltonian perturbation theory. More precisely, we show with a suitable rescaling of the variables that the system is equivalent to a small, smooth and adiabatic perturbation of a simple one-dimensional completely integrable system: a particle moving freely on a circle. We then use Lie-Deprit perturbation theory to show momentum is an adiabatic invariant of the system. The result actually holds for arbitrary time-dependent potentials, provided they are bounded and sufficiently smooth in time: no assumption needs to be made on the (quasi-)periodicity of the time-dependence.  Since the system is one-dimensional, we have no resonances to contend with. These results are not surprising: other one-dimensional non-autonomous systems have been studied extensively in the literature over the years, such as the Fermi-Ulam model \cite{ulam} in which a particle moves between a fixed and a periodically moving wall, and variations thereof. It is generally expected and it has been shown in some cases that sufficiently smooth, periodically driven one-dimensional systems are stable: the basic idea is that some KAM tori survive the perturbation and create barriers blocking indefinite energy growth \cite{lili1,lili2, lm}. Presumably, if one assumes the potential to be analytic, similar results can be obtained for the model we study here. 

It is finally instructive to notice the difference in behaviour of the pulsed rotors considered above with kicked rotors, in which the time dependence of the potentials in~\eqref{eq:H} is very singular, of the form
\begin{equation}\label{eq:kicked}
V(q,t)= \lambda\sum_n\ \delta(t-n)v(q),
\end{equation}
with $v$ a sufficiently smooth function of its argument.
For such systems, it is easy to write down the Floquet transformation which gives the evolution of the system over one period of the potential:
\begin{equation}\label{eq:kickfloquet}
\Phi(q,p)=(q', p'), \quad \mathrm{where}\quad p'=p-\lambda\nabla v(q),\ q'=q+p'.
\end{equation}
In that case, one finds for $\lambda$ sufficiently large, and {\em independently of the dimension $d$}, that (see Figure~\ref{fig:kickedd1})
\begin{equation}
\left< \|p(t)\|^2\right> \sim t, \qquad \left< \|q(t)\|^2\right>\sim t^3.
\end{equation} 
This is again easily understood in terms of the random walk picture we develop in Section \ref{s:rw}, showing the flexibility of the approach developed in \cite{adblafp} and here. The main difference with the case of pulsed systems resides in the observation that, whereas for kicked systems the momentum change undergone by a particle in one period of the potential is of order $\|p\|^0\sim1$, independently of the size of the initial momentum of the particle and of the dimension $d$ (as is clear from (\ref{eq:kickfloquet})), this momentum change is of order $\| p\|^{-1}$ for pulsed systems. Indeed, in one period, the particle traverses the unit cell $\|p\|$ times, undergoing each time a momentum change of order $1/\|p\|^2$ only, as pointed out above. This fully explains 
the slower energy growth observed in pulsed systems, as well as the slower growth of the mean squared displacement.

We end with some comments on the situation in quantum mechanics and the relation of our work with the question of quantum suppression of classical chaos. It was observed numerically in \cite{cciv} that the one-dimensional quantum kicked rotor is stable, even for large coupling ($\lambda>>1$), in contrast to its classical counterpart, which, as pointed out above, is stable only at small coupling.   This is the phenomenon that has been dubbed ``quantum suppression of classical chaos''; we refer to \cite{i} \cite{f} for reviews on the quantum kicked rotor in one dimension. Although stability of the kicked quantum rotor for {\em weak} coupling was proved in \cite{bo02}, the problem remains open for strong coupling, therefore leaving a rigorous proof of the occurrence of quantum suppression of classical chaos in this system open.  

For quantum pulsed rotors, on the other hand, the situation is somewhat better understood. First, in one dimension, it follows from \cite{dsvv} that if 
the potential is small, time-periodic with suitably Diophantine period, and sufficiently smooth in both its variables,  the system is indeed stable \cite{duclos}. The proof relies on a quantum version of KAM theory  and exploits the widening energy  gap between successive eigenvalues of the unperturbed quantum system, typical of the free particle on a one-dimensional torus. It is however not clear whether this is an example of ``quantum suppression of classical chaos'', to the extent that, as we argue here, the classical pulsed rotor is also stable in one dimension, and this independently of the strength of the potential.
 
For quantum pulsed rotors of arbitrary dimension, on the other hand, various almost stability results have been proved by Bourgain, who showed they have a very slow  energy growth bounded by $t^\epsilon$, for any $\epsilon>0$, provided $V$ is a bounded $C^\infty$ function of  both its variables (see \cite{bo1}), and this, whether it is time (quasi-)periodic or not (see also~\cite{del}). If the potential is analytic, small, and quasi-periodic in time, then in two dimensions this growth is even bounded by a suitable power of $\ln t$ under some additional Diophantine conditions on the frequency vector (see \cite{bo2}). The same logarithmic bound holds in one dimension, without the smallness hypothesis. It furthermore holds for time-periodic potentials in all dimensions, without a smallness condition or conditions on the frequency vector. 

These quantum ``almost'' stability results for pulsed rotors stand in contrast to our results here on their classical counterparts, which we argue to be unstable if $d\geq 2$, with a power law growth of their kinetic energy given by $t^{2/5}$. As such, these models will provide examples of ``quantum suppression of classical chaos'' provided a fully rigorous proof can be provided of our results on the higher dimensional systems considered in this paper.

The rest of the paper is organized as follows. In Section~\ref{s:rw} we describe the random walk approximation we use to analyze the motion of the systems studied when $d\geq 2.$ We also show how to derive the various power laws for $\left<\|p(t)\|^2\right>$ and $\left<\|q(t)\|^2\right>$ from it and present our numerical data. In Section~\ref{s:onedrotor}, we study the one dimensional problem. In Section~\ref{s:kicked}, we establish a comparison with kicked systems.

\noindent{\bf Acknowledgments:} SDB wishes to thank his regretted friend and colleague  Pierre~Duclos for explaining his recent work \cite{dsvv} to him and for illuminating discussions on the quantum stability problem. The authors also thank J.~C.~Alvarez, N.~Berglund and P.~Lochak for helpful discussions on Hamiltonian perturbation theory. This work was supported by the Ministry of Higher Education and Research,
Nord-Pas de Calais Regional Council and FEDER through the ``Contrat de
Projets \'Etat R\'egion (CPER) 2007-2013''

\begin{figure}[tbp]
\begin{center}
\psfrag{en}{$e_n$}
\psfrag{yn-}{$q_n^-$}
\psfrag{vn}{$p_n$}
\psfrag{bn}{$b_n$}
\psfrag{yn}{$x_{N_n}$}
\psfrag{1/2}{$1/2$}


\includegraphics[height=7cm,keepaspectratio=true]{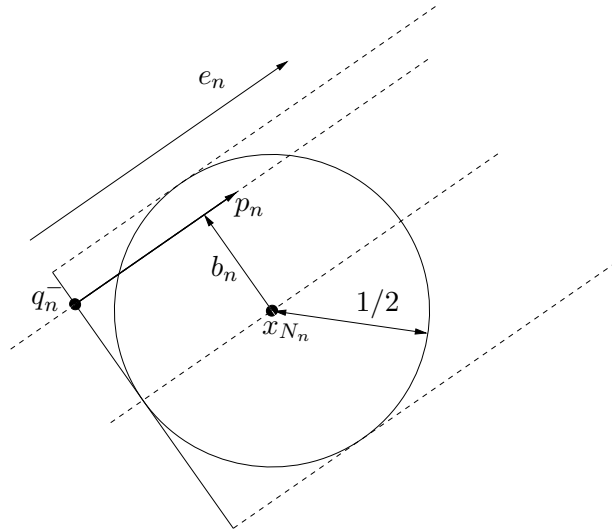}
\end{center}
\caption{A particle at time $t_n$ impinging with momentum $p_n$ and
impact parameter $b_n$ on the $n$th scatterer, centered at the
point $x_{N_n}.$}
\label{fig:collision}
\end{figure}

\section{Inelastic and elastic Lorentz gases in $d\geq 2$}\label{s:rw}

\begin{figure}
\begin{center}
\psfrag{-1.5}{{\tiny $-1.5$}}
\psfrag{-1.0}{{\tiny $-1$}}
\psfrag{-0.5}{{\tiny $-0.5$}}
\psfrag{0.0}{{\tiny $\ 0$}}
\psfrag{0.5}{{\tiny $0.5$}}
\psfrag{1.0}{{\tiny $1$}}
\psfrag{1.5}{{\tiny $1.5$}}
\psfrag{2.0}{{\tiny $2$}}
\psfrag{2.5}{{\tiny $2.5$}}
\psfrag{3.0}{{\tiny $3$}}
\psfrag{3.5}{{\tiny $3.5$}}
\psfrag{0}{{\tiny $0$}}
\psfrag{100}{{\tiny $100\ $}}
\psfrag{200}{{\tiny $200$ }}
\psfrag{300}{{\tiny $300$ }}
\psfrag{400}{{\tiny $400$ }}
\psfrag{500}{{\tiny $500$ }}
\psfrag{600}{{\tiny $600$ }}
\psfrag{5}{{\tiny $5$}}
\psfrag{10}{{\tiny $10$}}
\psfrag{15}{{\tiny $15$}}
\psfrag{20}{{\tiny $20$}}
\psfrag{25}{{\tiny $25$}}
\psfrag{30}{{\tiny $30$}}
\psfrag{35}{{\tiny $35$}}
\psfrag{40}{{\tiny $40$}}
\psfrag{45}{{\tiny $45$}}
\psfrag{50}{{\tiny $50$}}
\psfrag{1}{{\tiny $1$}}
\psfrag{2}{{\tiny $2$}}
\psfrag{3}{{\tiny $3$}}
\psfrag{4}{{\tiny $4$}}
\psfrag{6}{{\tiny $6$}}
\psfrag{7}{{\tiny $7$}}
\psfrag{-0.25}{{\tiny $-0.25$}}
\psfrag{0.25}{{\tiny $0.25$}}
\psfrag{-0.02}{{\tiny $-0.02$}}
\psfrag{-0.01}{{\tiny $-0.01$}}
\psfrag{0.00}{{\tiny $0$}}
\psfrag{0.01}{{\tiny $0.01$}}
\psfrag{0.02}{{\tiny $0.02$}}
\psfrag{0.03}{{\tiny $0.03$}}
\psfrag{0.04}{{\tiny $0.04$}}
\psfrag{0.05}{{\tiny $0.05$}}
\psfrag{0.06}{{\tiny $0.06$}}
\psfrag{0.07}{{\tiny $0.07$}}
\psfrag{k}{{\scriptsize$k$}}
\psfrag{nb2part}{{\scriptsize number of particles}}
\psfrag{b}{{\scriptsize $b_n$}}
\psfrag{phi}{{\scriptsize $\phi_n$}}
\psfrag{<b_n*b_{n+k}>-<b_n><b_{n+k}>}{{\scriptsize$b_{n,k}$}}
\psfrag{<phi_n*phi_{n+k}>-<phi_n><phi_{n+k}>}{{\scriptsize$\phi_{n,k}$}}
\psfrag{v8}{{\tiny $\left\|p_0\right\|=1.36$}}
\psfrag{v3}{{\tiny $\left\|p_0\right\|=0.82$}}
\psfrag{v0}{{\tiny $\left\|p_0\right\|=0.50$}}
\subfigure[Distribution of $b_{n}$]{\includegraphics[height=3.4cm,width=6cm]{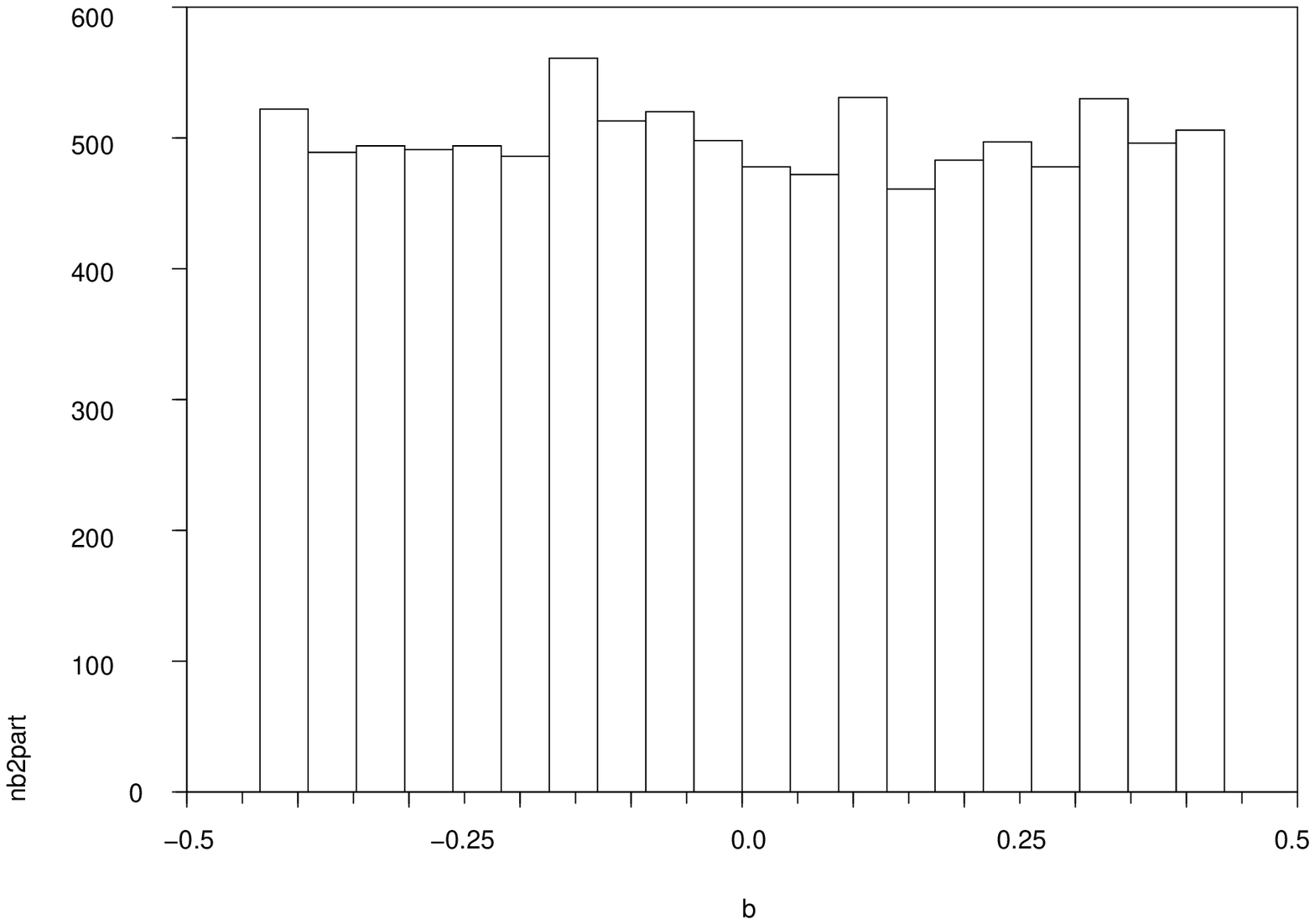}}
\subfigure[$b_{n,k}:=\left<b_nb_{n+k}\right>-\left<b_n\right>\left<b_{n+k}\right>$]{\includegraphics[height=3.4cm,width=6cm]{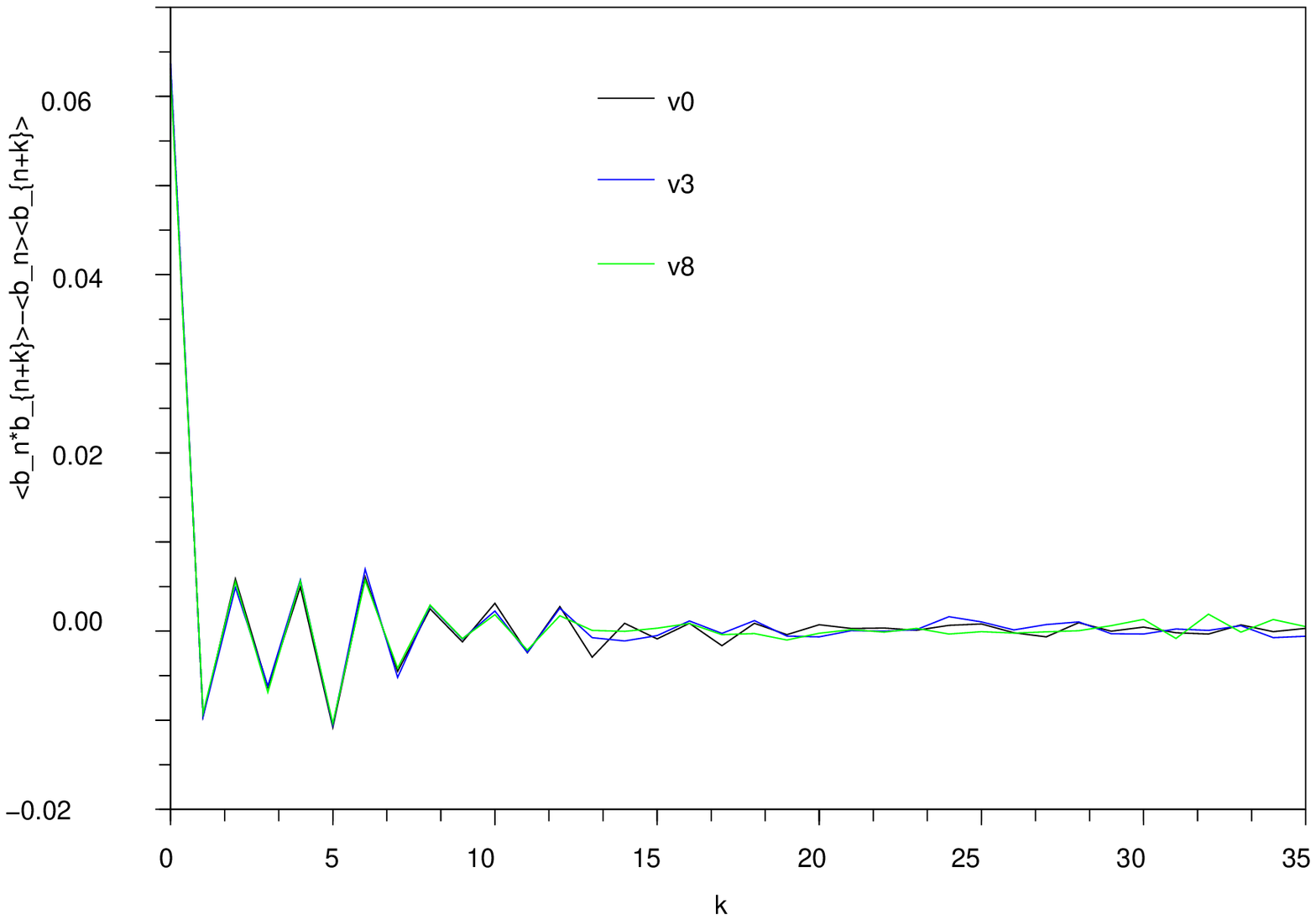}}
\subfigure[Distribution of $\phi_{n}$]{\includegraphics[height=3.4cm,width=6cm]{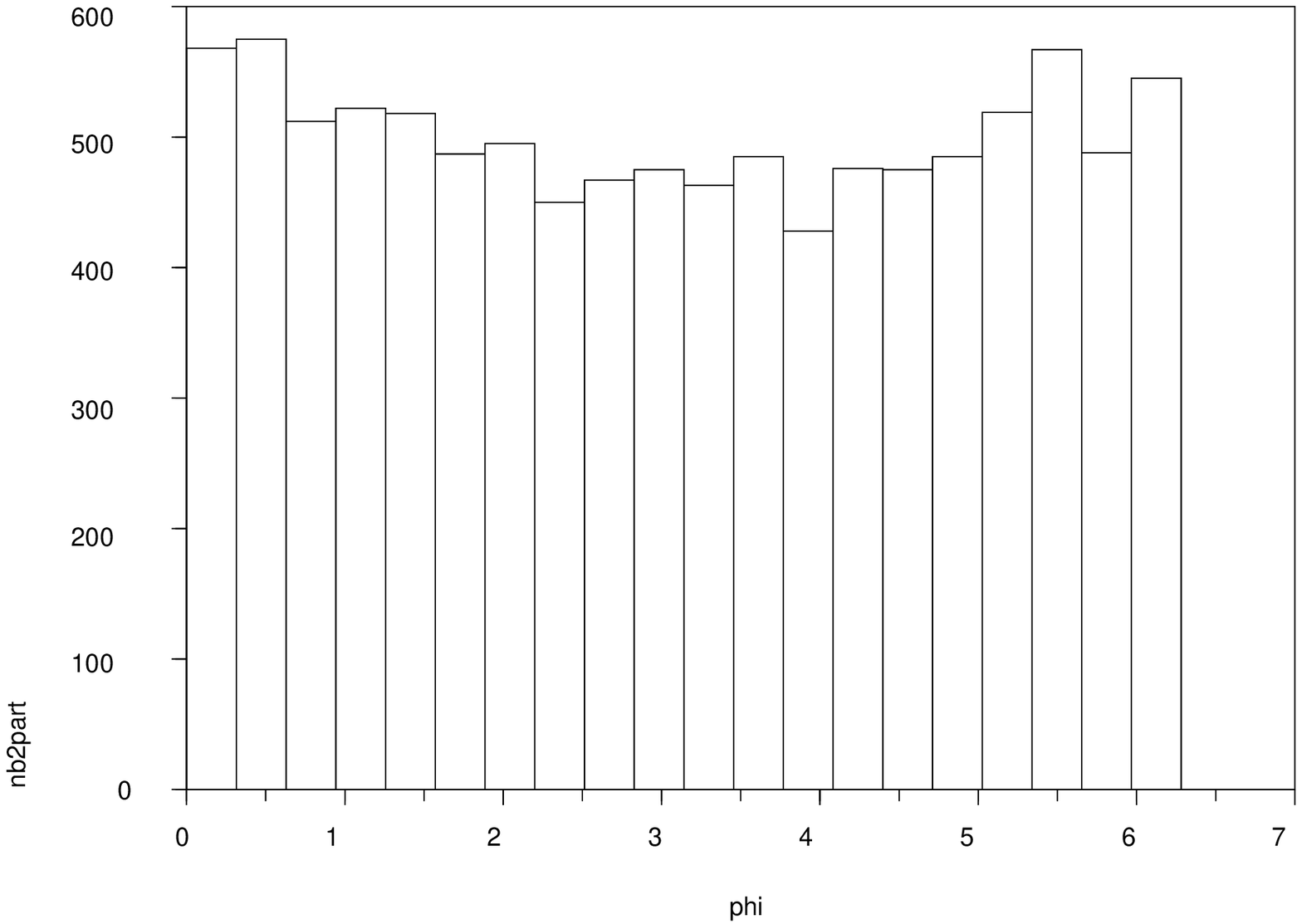}}
\subfigure[$\phi_{n,k}:=\left<\phi_n\phi_{n+k}\right>-\left<\phi_n\right>\left<\phi_{n+k}\right>$]{\includegraphics[height=3.4cm,width=6cm]{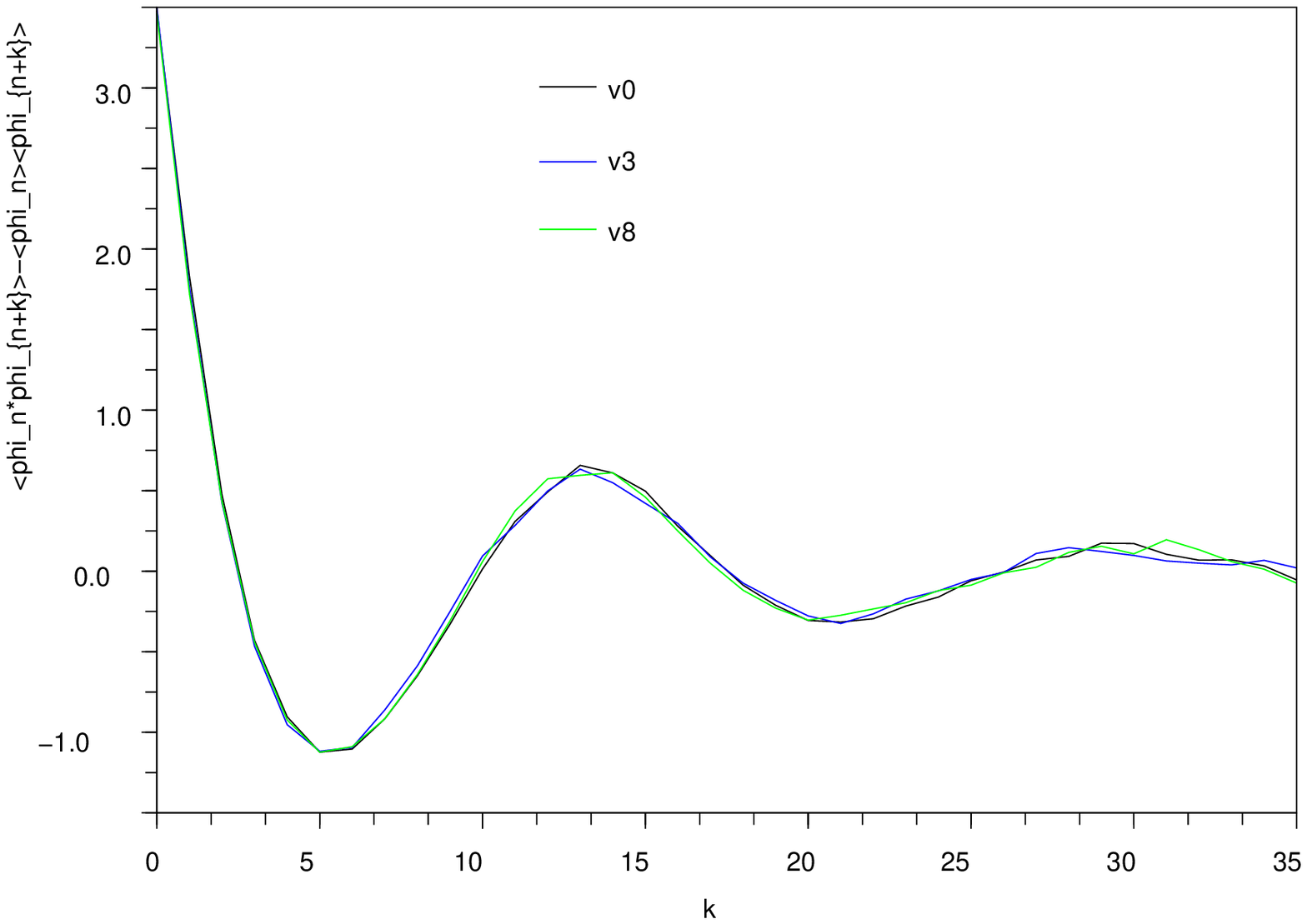}}
 \caption{Distributions and correlations of $b_n$ and $\phi_n,$ with $n=50000,$ for the model described by equation~(\ref{eq:numericsmodel}) where $f(t)=\cos(t),$ calculated with $10^4$ trajectories. \label{fig:d2correl}}
\end{center}
\end{figure}

In this section, we first briefly explain how the random walk description of the motion through an array of random scatterers, developed in \cite{adblafp}, can be adapted to the present case.  We suppose $d\geq 2.$ We will then summarily show how the analysis of this random walk leads to the asymptotics (\ref{eq:d2results}), referring to \cite{adblafp} for more details. We consider a typical trajectory of a particle moving in the potential $V$ described in (\ref{eq:V}).
Such a trajectory will successively ``visit'' scattering centers $x_{N_n}$ at a sequence of instants $t_n$, with incoming momenta $p_n$ and impact parameters $b_n$.  Those quantities are related by (see Figure~\ref{fig:collision}):
$$
q_n^-=x_{N_n}-\frac12 e_n + b_n, \quad e_n=\frac{p_n}{\| p_n\|}\quad b_n\cdot e_n=0,\quad \|b_n\|\leq 1/2.
$$ 
Note that we view the impact parameter as a vector perpendicular to the incoming direction. 
We choose $t_0=0$ and $x_0=0$: the particle starts at the scatterer at the origin. We also assume the initial velocity of the particle is large, meaning $\|p_0\|^2 >>\lambda$. At each scattering event the momentum of the particle undergoes a change $\Delta p_n= p_{n+1}-p_n$. The momentum $p_{n+1}$ with which it leaves the $n$-th scatterer is also the one with which it impinges on the next one. This momentum change depends on the time of arrival $t_n$ of the particle at the scattering center, via the phase $\phi_n=\phi_0+\omega t_n$, as well as on its impact parameter $b_n$. One has
\begin{equation}\label{eq:deltap}
\Delta p_n = R(p_n, b_n, \phi_n),
\end{equation}
where 
\begin{equation}\label{eq:R}
R(p, b, \phi)=-\lambda\int_0^{+\infty} \rd \tau  \nabla W(q(\tau), \omega\tau+\phi)
\end{equation}
in which $\tau\to q(\tau)$ is the solution of
$$
\ddot q(t)=-\lambda\nabla W(q(t), \omega t +\phi),\ q(0)=b-\frac12\frac{p}{\| p\|},\ p(0)=p.
$$
After leaving the support of the $n$th scatterer, the particle travels a distance $\eta_n$ to the next scatterer, which it reaches after a time $\Delta t_n=\eta_n/\| p_{n+1}\|$. The distance $\eta_n$, hence the time $t_{n+1}$ and the phase $\phi_{n+1}$ depend on the geometry of the periodic lattice as well as on the dynamics of the scattering event via the precise point $q_n^+$ where the particle leaves the $n$-th scatterer, as well as through the outgoing direction $e_{n+1}$. As a result, it is legitimate to expect that, after a large number of scattering events, the phases $\phi_n$ and impact parameters $b_n$ are randomized, that is, they are uniformly distributed in their natural range, and that they display short temporal correlations only. This hypothesis was tested numerically, and the results are shown in Figure~\ref{fig:d2correl}. 
In all the numerical results shown in this paper for higher dimensional systems, we employed a two-dimensional hexagonal lattice with, for $N=\left(N_1,N_2\right)\in\Z^2,\ x_N=N_1u+N_2v,$ where $u=(1,0),\ v=\frac12(1,\sqrt{3}).$ The potential $W$ was taken in the form of a time-dependent, flat circular potential,
\begin{equation}\label{eq:numericsmodel}
W(q,t)=f(t)\chi\left(\frac{\|q\|}{q_*}\right),\qquad q\in\R^2,
\end{equation}
where $\chi(x)=1$ if $0\leq x\leq 1$ and $\chi(x)=0$ otherwise. The parameter $q_*$ satisfies $\frac{\sqrt{3}}{4}< q_*<1/2$ to ensure the system has finite horizon and $\lambda=1/6$. Two different cases were explored for the function $f,$ namely,
$$f_1(t)=\cos(t)\qquad\textrm{and}\qquad f_2(t)=\cos(t)+\cos\left(\sqrt{2}t\right).$$
In the numerical calculations, each particle has the same initial speed and is initially placed at random at a point on the boundary of the scatterer at the origin, with an initial velocity drawn with equal probability from all possible outward directions.

The data in Figure~\ref{fig:d2correl} strongly indicate that the distributions of phases and impact parameters are indeed randomized after a relatively short time, and that they decorrelate, despite the fact that the potential is periodic in space and time.
Under the described circumstances, the statistical behaviour of the ensemble of particles is expected to be the same as that of an ensemble executing the random walk
\begin{eqnarray}\label{eq:rw}
p_{n+1}=p_n+ R(p_n,b_n, \phi_n), \ t_{n+1}=t_n + \frac{\eta_*}{\| p_{n+1}\|},  q_{n+1}= q_n + \eta_* e_{n+1},
\end{eqnarray}
where $\eta_*$ is the average distance traveled between scattering events, $q_n$ the position of the particle at the $n$th such event and in which $b_n$ and $\phi_n$ are treated as independent random variables, each distributed in its natural range. We will write $\langle \cdot \rangle$ for the corresponding averages.

The analysis of this random walk was performed in detail in \cite{adblafp} and in \cite{a}. We reproduce here only the essence of the arguments allowing to predict the asymptotic behaviour of the particles' averaged energy and position, referring to the above works for more mathematical details. We  will in particular only carry the leading order terms in all expansions.

\begin{figure}
\begin{center}
\psfrag{-1}{{\tiny $0.1$}}
\psfrag{0}{{\tiny $1$}}
\psfrag{1}{{\tiny $10$}}
\psfrag{2}{{\tiny $10^2$}}
\psfrag{3}{{\tiny $10^3$}}
\psfrag{4}{{\tiny $10^4$}}
\psfrag{5}{{\tiny $10^5$}}
\psfrag{6}{{\tiny $10^6$}}
\psfrag{7}{{\tiny $10^7$}}
\psfrag{8}{{\tiny $10^8$}}
\psfrag{9}{{\tiny $10^9$}}
\psfrag{10}{{\tiny $10^{10}$}}
\psfrag{12}{{\tiny $10^{12}$}}
\psfrag{v8}{{\tiny $\left\|p_0\right\|=1.36$}}
\psfrag{v3}{{\tiny $\left\|p_0\right\|=0.82$}}
\psfrag{v0}{{\tiny $\left\|p_0\right\|=0.50$}}
\psfrag{power1}{{\tiny $\sim t^{2/5}$}}
\psfrag{power2}{{\tiny $\sim n^{1/3}$}}
\psfrag{power3}{{\tiny $\sim t^2$}}
\psfrag{power4}{{\tiny $\sim n^{5/3}$}}
\psfrag{n}{{\scriptsize$n$}}
\psfrag{t}{{\scriptsize$t$}}
\psfrag{p_n}{{\scriptsize$\left<\|p_n\|^{2}\right>$}}
\psfrag{p_t}{{\scriptsize$\left<\|p(t)\|^{2}\right>$}}
\psfrag{q_n}{{\scriptsize$\left<\|q_n\|^2\right>$}}
\psfrag{q_t}{{\scriptsize$\left<\|q(t)\|^2\right>$}}
\includegraphics[height=7cm,keepaspectratio=true]{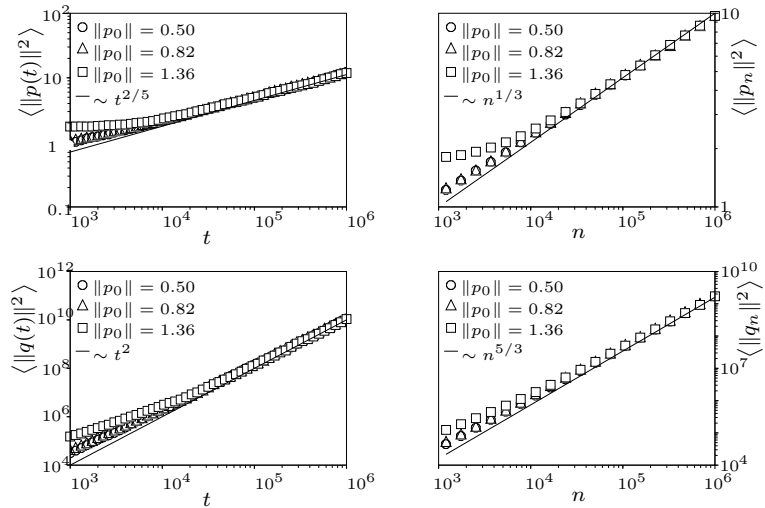}
\end{center}
 \caption{Numerically determined values of $\left<\|p_n\|^{2}\right>,$ $\left<\|p(t)\|^{2}\right>,$
$\left<\|q_n\|^2\right>$ and $\left<\|q(t)\|^2\right>$ in a two-dimensional hexagonal lattice for the potential described by equation~(\ref{eq:numericsmodel}) where $f(t)=\cos(t)$. In each plot, different symbols correspond to different initial conditions, as indicated.} \label{fig:d2pulsedp}
\end{figure}

We first describe the growth of the kinetic energy of the particle (see Figure~\ref{fig:d2pulsedp}, top panels). Leading order perturbation theory applied to (\ref{eq:deltap}) and (\ref{eq:R}) yields 
\begin{equation}\label{eq:normprw}
\| p_{n+1}\| \simeq \| p_n\| + \frac{\beta_n^{(1)}}{\| p_n\|^2},
\ 
\mathrm{where}\ 
\beta_n^{(1)}=\lambda\int_{-\infty}^{+\infty}\rd \mu\ \left(\omega\cdot\nabla_\phi\right) W\left(b+\mu e_n,\phi_n\right).
\end{equation}
Since, if the $\phi_n$ are uniformly distributed on the torus, $\langle\beta_n^{(1)}\rangle=0$, one can think of (\ref{eq:normprw}) as a random walk with variable step size and it is then easy to determine the asymptotic behaviour of $\| p_n\|$. Indeed, since
\begin{equation}\label{eq:missing1}
\Delta \|p_n\|^3 \simeq 3\|p_n\|^2\Delta \| p_n\| \simeq 3\beta_n^{(1)},
\end{equation}
$\| p_n\|^3$ executes a standard random walk with $\|p_n\|$-independent step, so that
\begin{equation}\label{eq:missing2}
\left<\| p_n\|^3\right>\sim n^{\frac12},\quad \mathrm{and}\quad \left< \|p_n\|^s\right>\sim n^{\frac{s}6}\ (s>-3),
\end{equation}
since the correlations between the $\beta_n^{(1)}$ are supposed to decrease fast.
Combining this result with (\ref{eq:rw}), one finds
\begin{equation}\label{eq:ptd2}
\left< t_n\right> \sim n^{\frac{5}{6}}, \quad \mathrm{so\ that}\quad \left< \| p(t)\|^s\right> \sim t^{\frac{s}{5}},
\end{equation}
which proves the first equation in (\ref{eq:d2results}). This slow growth of the energy ($s=2$) is indeed observed in the numerical data presented in Figure~\ref{fig:d2pulsedp} (top panels). It is identical to the one found for random time-dependent potentials in \cite{adblafp} confirming that the motion in a time-periodic potential of the type considered is self-randomizing due to the instabilities inherent in the scattering mechanism.

To understand the asymptotic behaviour of the particle's position $q(t)$ we have to analyze how fast it turns. For that purpose, notice first that (\ref{eq:R}) and (\ref{eq:rw}) allow one to write, to lowest order in perturbation theory:
\begin{equation}\label{eq:prw}
p_{n+1}\simeq p_n + \frac{\alpha^{(1)}_n}{\| p_n\|},
\ \mathrm{where}\ 
\alpha_n^{(1)}=-\lambda\int_{-\infty}^{+\infty} \rd \mu \ \nabla W\left(b_n+\mu e_n,\phi_n\right).
\end{equation}
Note that $\alpha_n^{(1)}\cdot e_n=0$ and $\langle\alpha_n^{(1)}\rangle=0$. It follows from (\ref{eq:normprw}) and (\ref{eq:prw}) that
\begin{equation}\label{eq:directionalwalk}
e_{n+1}\simeq e_n + \frac{\alpha_n^{(1)}}{\| p_n\|^2}.
\end{equation}
One can therefore think of the $e_n$ as executing a random walk on the unit sphere in $\R^d$.  For values of $m$ small enough that $\| p_{n+m}\|\sim \| p_n\|$, that is for $m<<n$ (see (\ref{eq:missing1})-(\ref{eq:missing2})), it is legitimate to approximate this walk by
$$
e_{n+m}\simeq e_n + \frac1{n^{\frac13}}\sum_{k=0}^{m-1}\alpha_{n+k}^{(1)}.
$$
Thinking of the $\alpha_{n+k}^{(1)}$  as independent steps (again, for a more careful treatment, see again \cite{adblafp}), it  follows that
$
\left<\|e_{n+m}-e_n\|\right>\sim{m^{\frac12}}/{n^{\frac13}}.
$
\begin{figure}
\begin{center}
\subfigure[$m=20$]{\includegraphics[height=3cm,keepaspectratio=true]{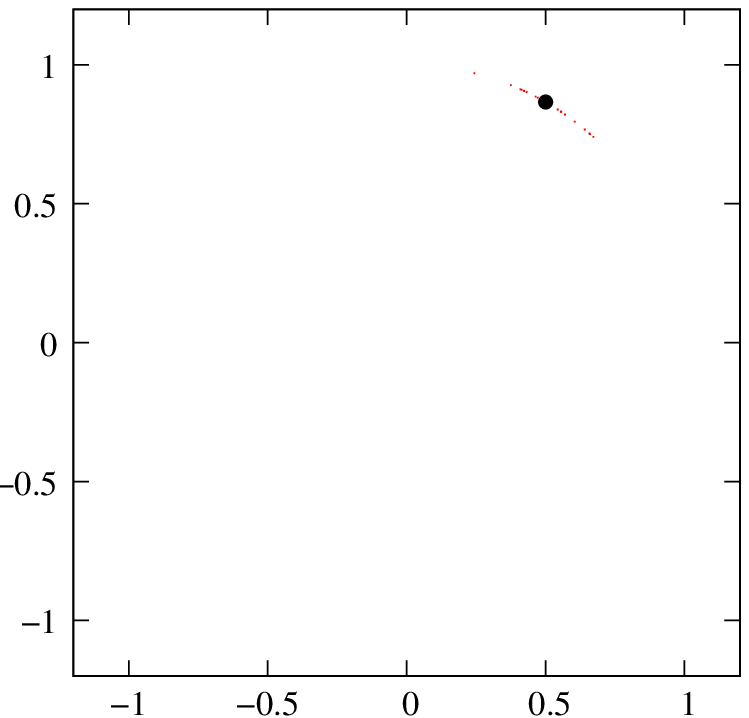}}
\subfigure[$m=100$]{\includegraphics[height=3cm,keepaspectratio=true]{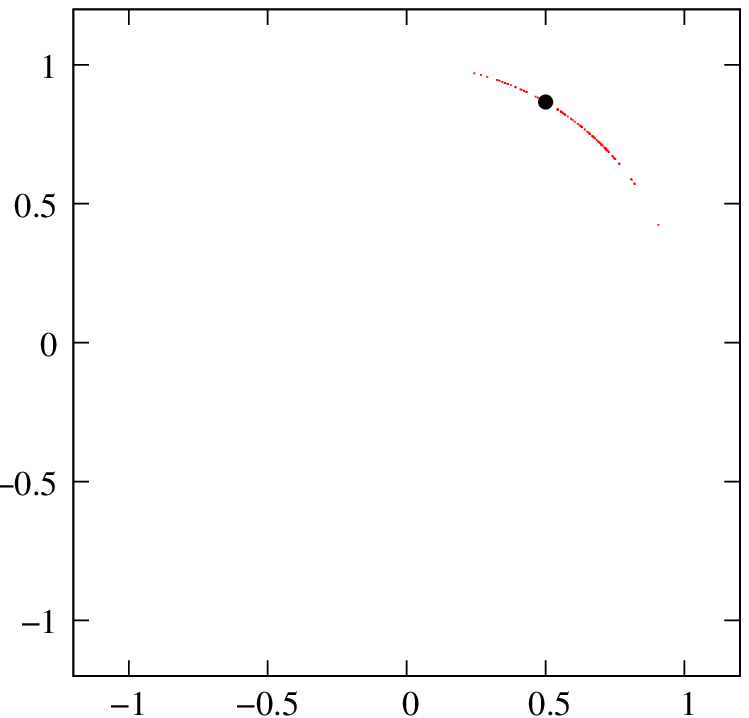}}
\subfigure[$m=700$]{\includegraphics[height=3cm,keepaspectratio=true]{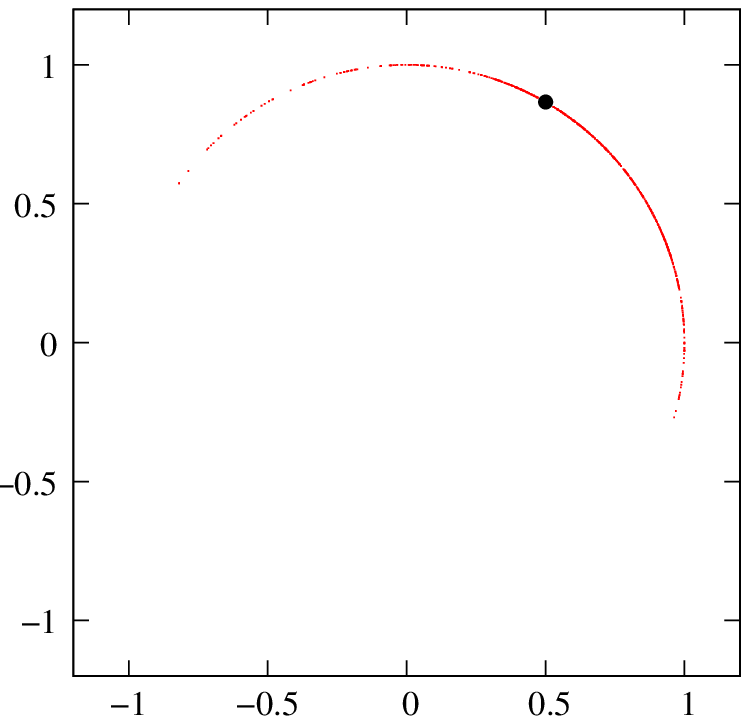}}
\subfigure[$m=1500$]{\includegraphics[height=3cm,keepaspectratio=true]{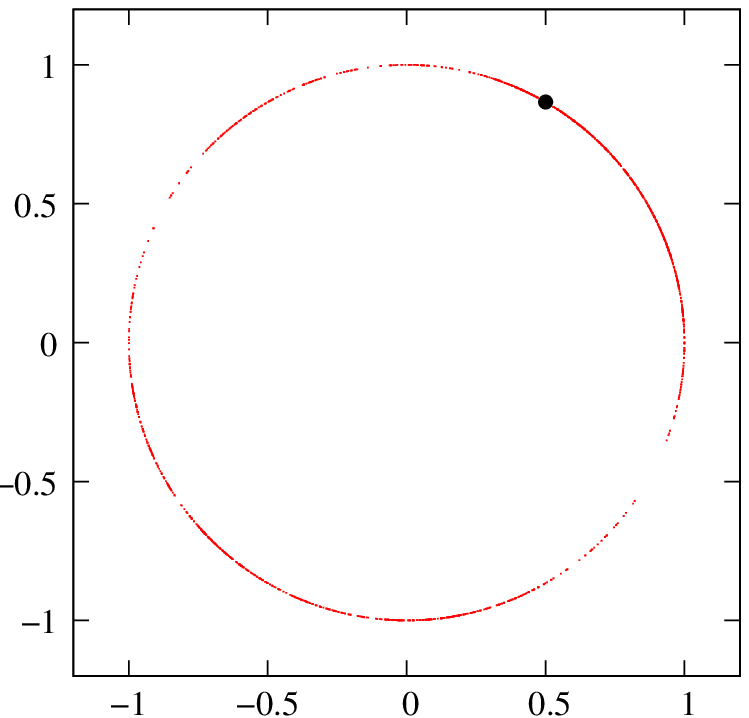}}
 \caption{Coordinates of the vectors $e_k,$ $k\in[\![1;m]\!],$ with $m$ as indicated, for a single trajectory in the model described by equation~(\ref{eq:numericsmodel}) with $f(t)=\cos(t).$ The initial velocity is taken to be $p_0=2e_0,$  and $e_0$ is represented by the large dot.}
\label{fig:d2pulseden}
\end{center}
\end{figure}
We conclude that $\left<\|e_{n+m}-e_n\|\right>\sim 1$ for times of order $m\sim n^{\frac23}<<n$.
This means in particular that, if after $n$ collisions, we consider only those particles moving in a given direction $e,$ so that $e_n=e$, it will take of the order of $n^{2/3}$ more collisions for the  distribution of  particle directions to spread over a macroscopic angle about $e$.  This spreading is qualitatively illustrated in Figure~\ref{fig:d2pulseden}, where we see that after $m\sim 1500$ collisions, the set of directions $\left\{e_k\right\}_{k\in[\![1;m]\!]}$ covers the unit circle.

Writing $M_k$ for the number of collisions needed for the particle to turn $k$ times over a macroscopic angle, we find
$$
M_{k+1}=M_k+M_k^{\frac23},
$$
so that $M_k\sim k^3$. It is then reasonable to think of the motion as being effectively rectilinear at constant speed between $M_k$ and $M_{k+1}$, so that  one obtains an effective random walk for the particle position given by
$$
q_{M_{k+1}}=q_{M_k}+\eta_*(M_{k+1}-M_k)e_{M_k}\simeq q_{M_k}+\eta_* k^2 e_{M_k}.
$$
Treating the $e_{M_k}$ as independent random variables distributed uniformly on the sphere, one finds from this that
$$
\left< \|q_{M_k}\|^2\right>\sim \sum_{\ell=1}^k \ell^4\sim k^5\simeq M_k^{\frac53}.
$$
Extrapolating to all $n$, and using (\ref{eq:ptd2}) yields the second equation in (\ref{eq:d2results}) (see Figure~\ref{fig:d2pulsedp}, bottom panels).

We point out that the previous analysis about the evolution of $q_n$ applies equally well to a time-independent potential $W$. In that case, the kinetic energy $\left\|p_n\right\|^2$ of the particle is of course a constant of the motion and one finds, instead of (\ref{eq:directionalwalk})
$$
e_{n+1}\simeq e_n + \frac{\alpha_n^{(1)}}{\| p_0\|^2},
$$
so that 
$$
e_{n+m}=e_n+\frac{1}{\| p_0\|^2}\sum_{k=0}^{m-1}\alpha_{n+k}^{(1)}, \ \mathrm{and}\ \left\langle \| e_{n+m}-e_n\|\right\rangle\sim \frac{\lambda}{\|p_0\|^2}\sqrt{m},
$$
provided $\|p_0\|$ is large, that is $\|p_0\|^2>>\lambda$. The directions $e_n$ now diffuse on the unit sphere at a fixed rate, that decreases with the particle's speed as $\|p_0\|^{-2}$: indeed, the faster the particle, the less it turns in one scattering event. It follows that $M_{k+1}=M_k+\lambda^{-2}\|p_0\|^4,$ so that $M_k\sim\lambda^{-2}\parallel p_0\parallel^4 k$ from which one concludes with a reasoning as above that
$$
\left\|q(t)\right\|^2\sim \frac{\eta_*}{\lambda^2}\parallel p_0\parallel^5t.
$$  
\begin{figure}
\begin{center}
\psfrag{-1}{{\tiny $0.1$}}
\psfrag{0}{{\tiny $1$}}
\psfrag{1}{{\tiny $10$}}
\psfrag{2}{{\tiny $10^2$}}
\psfrag{3}{{\tiny $10^3$}}
\psfrag{4}{{\tiny $10^4$}}
\psfrag{5}{{\tiny $10^5$}}
\psfrag{6}{{\tiny $10^6$}}
\psfrag{8}{{\tiny $10^8$}}
\psfrag{10}{{\tiny $10^{10}$}}
\psfrag{12}{{\tiny $10^{12}$}}
\psfrag{v0^5}{{\tiny $\sim\left\|p_0\right\|^5$}}
\psfrag{power}{{\tiny $\sim t$}}
\psfrag{coef2diff}{{\tiny $<\|q(t)\|^2>/t$}}
\psfrag{power3}{{\tiny $\sim t^2$}}
\psfrag{power4}{{\tiny $\sim n^{5/3}$}}
\psfrag{t}{{\scriptsize$t$}}
\psfrag{q(t)^2}{{\scriptsize$\left<\|q(t)\|^2\right>$}}
\psfrag{v0}{{\scriptsize$\|p_0\|$}}

\includegraphics[height=7cm,keepaspectratio=true]{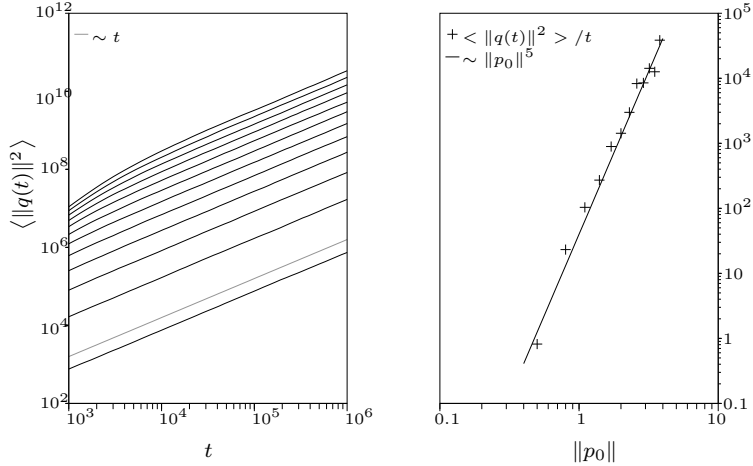}
\end{center}
\caption{Numerically determined values of $\left<\|q(t)\|^2\right>$ (left panel)  and of the diffusion constant (right panel) in the case of a soft Lorentz gas with a  potential as in (\ref{eq:numericsmodel}), with $f(t)=1$ and $\lambda=0.49^2/2,$ for $12$ initial velocities $\|p(0)\|$ from $0.5$ to $3.8$. \label{fig:softlorentz}}
\end{figure}
In other words, our random walk analysis shows that in a soft {\em elastic} Lorentz gas, the particle diffuses, and that the corresponding diffusion constant grows like $\left\|p_0\right\|^5$. This is illustrated in Figure~\ref{fig:softlorentz}. One notices in the left panel that at high speeds the diffusive regime sets in at later times, as expected: this is apparent from the bending of the topmost graphs. For short times, the motion is ballistic, of course. It is furthermore interesting to note the difference with what happens in the hard Lorentz gas, where diffusion of the particle has been rigorously proven; one has $\left\|q(t)\right\|^2\sim \|p_0\| t.$ The slow growth of the diffusion constant as $\|p_0\|$ can in that case be explained by the observation that, in the presence of hard obstacles, the 
trajectories do not depend on the speed of the particle, so that the velocity dependence of the diffusion constant follows from a simple scaling argument.

This ends our analysis of the class of soft non-dissipative inelastic and elastic Lorentz gases in dimension $d\geq 2$, described by (\ref{eq:V}).  As pointed out in the introduction, the Hamiltonians (\ref{eq:H})-(\ref{eq:V}) can also be viewed as describing pulsed rotors: our results here show they are unstable in higher dimension since their energy grows on average. A comparison with kicked rotors will be made in Section~\ref{s:kicked}.  But we first turn in the next section to the study of the one-dimensional situation, which is very different. 

\section{Inelastic Lorentz gases and pulsed rotors in one dimension}\label{s:onedrotor}
Unlike the standard (hard) Lorentz gas, which has no interesting analog in one dimension, the soft Lorentz gas described by (\ref{eq:H})-(\ref{eq:V}) yields a meaningful and non-trivial dynamics in one dimension as well. It is to its study we turn our attention in this section.
 
We have just shown that, in dimensions strictly higher than one, the asymptotic behaviour of a high speed particle in a potential periodic in space and (quasi-)periodic in time (as in (\ref{eq:V})) is identical to its behaviour in a field of random scatterers (as in (\ref{eq:randomV})). We will now see that in one dimension the situation is totally different. It was shown in \cite{adblafp} that in the random case, the particle momentum grows as $t^{1/5}$, just as in higher dimensions.   We will show here that, on the contrary, the momentum of a particle in a space and time periodic potential remains bounded for very long times. The precise results are given in Theorems~\ref{th:averaging} and~\ref{co:approxp} below. In fact, our conclusions hold for spatially periodic potentials with an arbitrary smooth and bounded time-dependence: time-periodicity is not needed. The results follow from time-dependent Hamiltonian perturbation theory, with the inverse of the particle's initial 
 speed as the small parameter.  

To understand the notion of high speed in the present context it is helpful to remember how the dimensionless Hamiltonian $H$, position $q$, momentum $p$ and time $t$  in (\ref{eq:H}) arise. The problems we are considering have a natural length scale $\ell$, which is the size of the unit cell of the lattice, and a natural time scale $T$, which is the typical scale on which the potential fluctuates. This suggests expressing the position $\tilde{q}$ of the particle  as a multiple of $\ell$, its momentum as a multiple of $\ell/T$ (we set the mass equal to $1$) and to measure time in units of $T$. So, writing $\tilde{q}=\ell q$,  $\tilde{p}=p\ell/T$ and $\tau=tT$, the equations of motion for $(\tilde{q},\tilde{p})$, equivalent to those obtained from (\ref{eq:H}) are
\begin{equation}  \label{eq:fastparticle}
\left.
\begin{split}
\frac{\rd \tilde{q}}{\rd \tau}(\tau) &=\tilde{p}(\tau),\\
\frac{\rd \tilde{p}}{\rd \tau}(\tau) &= -\frac{\lambda\ell}{T^2} \partial_q V\left(\frac{\tilde{q}(t)}{\ell}, \frac{\tau}{T}\right),
\end{split}%
\right\}
\end{equation}
which are Hamilton's equations for 
$$
\widehat H(\tilde{q},\tilde{p},\tau)=\frac{\tilde{p}^2}{2}+\mu V\left(\frac{\tilde{q}}{\ell},\frac{\tau}{T}\right),
$$
with $\mu=\lambda\ell^2/T^2$ having the dimension of an energy. In what follows, the potential $V$ is a dimensionless smooth function in $C^\infty(\R^2)$, periodic of period $1$ in its first (spatial) variable, and bounded with uniformly bounded derivatives in both its variables. It need not be of the specific form given in (\ref{eq:V}) and in particular, we do not assume the potential is (quasi-)periodic in time. As before, the spatial periodicity of the potential implies we can equally well think of the equations as describing a particle moving on a circle of circumference $\ell$ in a bounded time-dependent potential. 

This Hamiltonian has one dimensionless parameter, which is $\lambda={\mu T^2}/{\ell^2}$, where $\mu$ is the typical strength of the potential and $\ell^2/T^2$ the kinetic energy of a particle crossing the unit cell in a time $T$. We will study the solutions of the equations of motion for large initial momentum $\tilde{p}(0)$, by which we mean
$$
ap_*\frac{\ell}{T}\leq \tilde{p}(0)\leq Ap_*\frac{\ell}{T}\Leftrightarrow a p_*\leq p(0)\leq Ap_*<+\infty
$$
for some fixed $2\leq a\leq A<+\infty$ and for a large value of $p_*$, to be specified below; $p_*$ is therefore a second dimensionless parameter, measuring the relative size of the initial momentum ${\tilde p}(0)$ and $\ell/T$.  Note that
$
\varepsilon=1/p_*
$
is the time a free particle with initial momentum $p_*$ needs to traverse the unit cell of length $1$, so that $\varepsilon<<1$  means this time is short compared to the time interval $1$, characteristic of the variations of the potential. The particle is in this sense fast. We will in addition suppose $\lambda \varepsilon^2<<1$, which means that the kinetic energy of the particle is much larger than the typical size of the potential: so the particle is energetic.

\begin{figure}
\begin{center}
\psfrag{-2}{{\tiny $0.01$}}
\psfrag{-1}{{\tiny $0.1$}}
\psfrag{0}{{\tiny $1$}}
\psfrag{1}{{\tiny $10$}}
\psfrag{3}{{\tiny $10^3$}}
\psfrag{4}{{\tiny $10^4$}}
\psfrag{5}{{\tiny $10^5$}}
\psfrag{6}{{\tiny $10^6$}}
\psfrag{7}{{\tiny $10^7$}}
\psfrag{8}{{\tiny $10^8$}}
\psfrag{9}{{\tiny $10^9$}}
\psfrag{11}{{\tiny $10^{11}$}}
\psfrag{13}{{\tiny $10^{13}$}}
\psfrag{14}{{\tiny $10^{14}$}}
\psfrag{v8}{{\tiny $\left\|p_0\right\|=1.36$}}
\psfrag{v3}{{\tiny $\left\|p_0\right\|=0.82$}}
\psfrag{v0}{{\tiny $\left\|p_0\right\|=0.50$}}
\psfrag{v1}{{\tiny $\left\|p_0\right\|=2$}}
\psfrag{v2}{{\tiny $\left\|p_0\right\|=3$}}
\psfrag{power3}{{\tiny $\sim t^2$}}
\psfrag{n}{{\scriptsize$n$}}
\psfrag{t}{{\scriptsize$t$}}
\psfrag{q_t}{{\scriptsize$q(t)^2$}}
\psfrag{p_t}{{\scriptsize$p(t)^2$}}


\includegraphics[height=5cm,width=11cm]{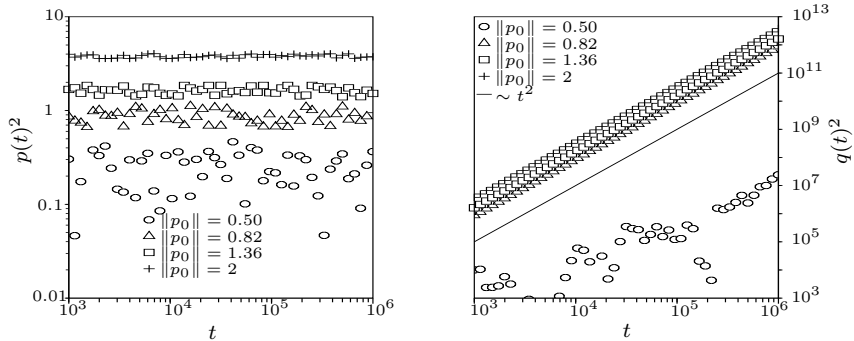}
\end{center}
\caption{Numerically determined values of  $p(t)^2$ and $q(t)^2$ in one dimension for the potential $V(q,t)=\cos t\chi_{[0,2/3]}(q)$. In each plot, different symbols correspond to different initial conditions, as indicated.}\label{fig:pulsedd1}
\end{figure}

What will happen is intuitively clear: since the particle is energetic and fast, the potential provides a small and slow perturbation of the free motion. In fact, in one traversal of the unit cell, the particle will pick up a momentum change of order $p_*^{-1}$ at most (see~(\ref{eq:prw})). Naively, one may expect that after a time of order $1$, the particle has turned around the circle $p_*$ times, resulting in a momentum change of order $1$. If the momentum changes were systematic, this could lead to an acceleration $p(t)\sim t$, much faster than in a random time dependent potential where it was proven in~\cite{adblafp} that $p(t)\sim t^{1/5}$. As we shall prove, the situation is much better than that: due to cancellations, the momentum is bounded for arbitrarily long times (Theorem~\ref{th:averaging}). The phenomenon is illustrated in Figure~\ref{fig:pulsedd1} for the potential $V(q,t)=\cos t \chi_{[0,2/3]}(q)$. Note that this potential is not even smooth in $q$, which doe
 s not seem to affect the stability of the momentum at all, on the time-scale explored. 
The difference with the situation in higher dimensions is intuitively clear: in one dimension, there are no impact parameters that can be randomized by the successive scattering events and in addition, since the distances between successive scattering events are all rigorously identical, the phases of the potential do not randomize either. 

\begin{figure}
\psfrag{-1.5}{{\tiny $ $}}
\psfrag{-1.0}{{\tiny $-1$}}
\psfrag{-0.5}{{\tiny $ $}}
\psfrag{0.0}{{\tiny $0$}}
\psfrag{0.5}{{\tiny $ $}}
\psfrag{1.0}{{\tiny $1$}}
\psfrag{1.5}{{\tiny $ $}}
\psfrag{2.0}{{\tiny $2$}}
\psfrag{2.5}{{\tiny $ $}}
\psfrag{0}{{\tiny $0$}}
\psfrag{20}{{\tiny $ $}}
\psfrag{40}{{\tiny $40$}}
\psfrag{60}{{\tiny $ $}}
\psfrag{80}{{\tiny $80$}}
\psfrag{100}{{\tiny $ $}}
\psfrag{120}{{\tiny $120$}}
\psfrag{140}{{\tiny $ $}}
\psfrag{160}{{\tiny $160$}}
\psfrag{180}{{\tiny $ $}}
\psfrag{200}{{\tiny $200$}}
\psfrag{1.88}{{\tiny $1.88$}}
\psfrag{1.90}{{\tiny $1.90$}}
\psfrag{1.92}{{\tiny $1.92$}}
\psfrag{1.94}{{\tiny $1.94$}}
\psfrag{1.96}{{\tiny $1.96$}}
\psfrag{1.98}{{\tiny $1.98$}}
\psfrag{2.00}{{\tiny $2$}}
\psfrag{2.02}{{\tiny $2.02$}}
\psfrag{0}{{\tiny $0$}}
\psfrag{2}{{\tiny $2$}}
\psfrag{4}{{\tiny $4$}}
\psfrag{6}{{\tiny $6$}}
\psfrag{8}{{\tiny $8$}}
\psfrag{10}{{\tiny $10$}}
\psfrag{12}{{\tiny $12$}}
\psfrag{14}{{\tiny $14$}}
\psfrag{16}{{\tiny $16$}}
\psfrag{18}{{\tiny $18$}}
\psfrag{20}{{\tiny $20$}}
\psfrag{p_0}{}
\psfrag{t}{{\scriptsize$t$}}
\psfrag{p_t}{{\scriptsize$p(t)$}}
\psfrag{p}{{\scriptsize$p\left(t\right)$}}
\psfrag{pav}{{\scriptsize$p_{ap}\left(t\right)$}}
\psfrag{tn}{{\scriptsize$t$}}
\hspace*{-1.cm}
\subfigure[Evolution of $p(t)$.]{\includegraphics[height=5cm,keepaspectratio=true]{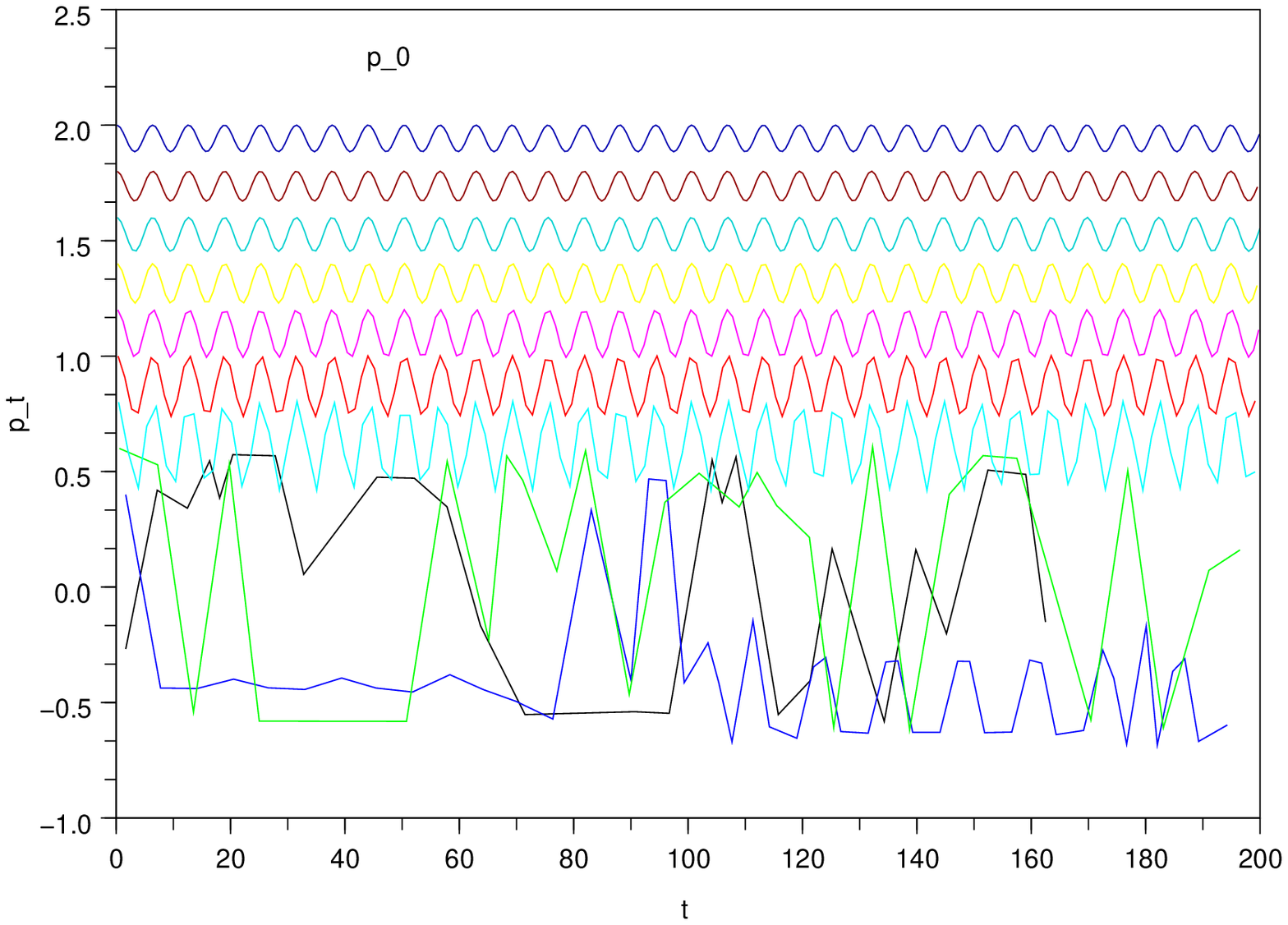}}
\hfill
\subfigure[ $p(t)$ vs $p_{ap}(t)$.]{\includegraphics[height=5cm,keepaspectratio=true]{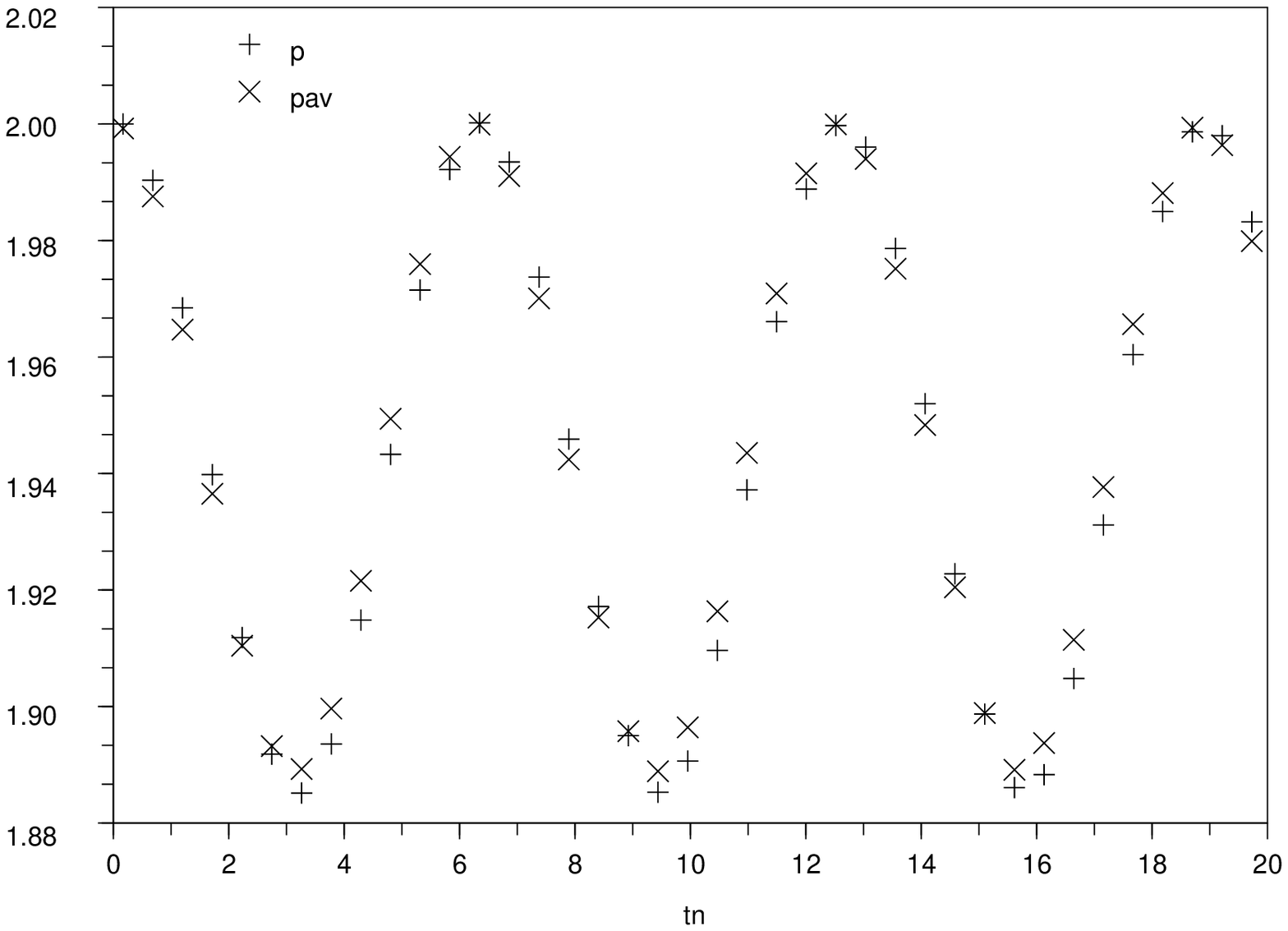}}
\caption{On the left, the evolution of $p(t)$ for ten different initial velocities, from $0.2$ to $2,$ for the potential $V(q,t)=\cos t \chi_{[0,2/3]}(q)$. On the right, numerically determined values of $p(t)$ compared to values of $p_{ap}(t)$ for the same model and with initial velocity $p(0)=2.$}\label{fig:d1pulsed}
\end{figure}

We now give  precise statements of the above claims, supposing
\begin{equation}\label{eq:hyppotential}
 \forall t\in\R, \int_0^1 V(q, t)\ \rd q=0.
\end{equation}
This condition constitutes no restriction since we can always substract its spatial average from the potential without changing the equations of motion. 
\begin{theorem}\label{th:averaging} Let $2\leq a\leq A<+\infty$. For all $\sigma_*>0$ and all $K\in\N^*$, there exists a constant $C_K>0$ such that, for all $p_*$ large enough and for all initial conditions
$
ap_*\leq p(0)\leq  A p_*, q(0)\in [0,1[,
$
one has, for $0\leq t\leq \sigma_*p_*^{K-2}$,
$$
| p(t)-p(0)|\leq C_K\frac1{p(0)}.
$$
\end{theorem}
The error estimate is optimal in this result, as  Theorem~\ref{co:approxp} below shows. The variations in $p(t)$ do really have an amplitude of $p(0)^{-1}$:
\begin{theorem}\label{co:approxp}
For all $\sigma_*>0$ and all $0\leq t\leq\sigma_*,$
$
p(t)=p_{ap}(t)+\rO\left(p_*^{-2}\right),
$
where
$
p_{ap}(t)=p(0)-{\lambda}\left(V\left(q(0)+p(0)t,t\right)-V\left(q(0),0\right)\right)/{p(0)}.
$
\end{theorem}
This result is illustrated in Figure~\ref{fig:d1pulsed}. Its proof provides in addition an approximation of $p(t)$ to order $p_*^{-2}$ for arbitrary long times (See Corollary~\ref{co:thelastone_yeahbaby!}).
We will prove these results using standard techniques of Hamiltonian perturbation theory as explained in~\cite{cary,lili3,lm}, only sketching the essence of the arguments and stressing the points particular to the present model. Details are given in \cite{a}. We note that Theorem~\ref{th:averaging} is close to a statement that can be found on p200 of ~\cite{lm}.
We first rewrite the equations of motion in a form suitable for Hamiltonian perturbation theory. For that purpose, we define a new time-scale and new variables
\begin{equation}\label{eq:defphiI}
\sigma=\frac{t}{\varepsilon}=p_*t\quad\bar{q}(\sigma)={q(\varepsilon \sigma)},\quad \bar{p}(\sigma)=\frac{p(\varepsilon \sigma)}{p_*}=\varepsilon p(\varepsilon \sigma).
\end{equation}
Here $\sigma$ counts the number of revolutions executed by a particle of momentum $p_*$ in a time  $t$. Then,  if $(q(t),p(t))$ solves Hamilton's equations corresponding to~\eqref{eq:H}, they satisfy
\begin{equation}\label{eq:standardform}
\left.\begin{split}
\frac{\rd \bar{q}}{\rd \sigma}(\sigma)&=\bar{p}(\sigma),\\
\frac{\rd \bar{p}}{\rd \sigma}(\sigma) &=-\varepsilon^2\lambda \partial_1 V(\bar{q}(\sigma), \varepsilon\sigma),\\
\end{split}\right\}
\end{equation}
with initial conditions $\bar{q}(0)\in[0,1[$ et $2\leq a\leq \bar{p}(0)\leq A<+\infty$. These are Hamilton's equations corresponding to $\overline{H}$ defined by
\begin{equation}\label{eq:defHbar}
\overline{H}(\bar{q},\bar{p},\varepsilon,\tau)=\frac12\bar{p}^2+\varepsilon^2\lambda V(\bar{q},\tau),
\end{equation}
where $\tau=\varepsilon\sigma.$ We will write 
$(\bar{q}(\sigma),\bar{p}(\sigma)):=\Phi_{\sigma}^{\overline{H}}(\bar{q}(0),\bar{p}(0))$
for the Hamiltonian flow generated by $\overline{H}.$ 

\noindent{\em Proof of Theorem~\ref{th:averaging}.} 
One considers a $\mathcal{C}^\infty$ function  $\chi(\bar{q},\bar{p},\mu,\tau)$ periodic in ${\bar q}$, defined for $\bar{p}>0,\mu,\tau\in\R$, which is  bounded together with all its derivatives on $\bar{p}>\delta$, for all $\delta>0$. We will show a judicious choice of $\chi$ allows to use its flow to define new coordinates  $(\overline{Q},\overline{P})$ in which the equations~\eqref{eq:standardform} take a simple form. More precisely, we define the local diffeomorphisms
$$
\Phi_{(\mu,\tau)}^\chi(\overline{Q},\overline{P})=(\bar{q}(\mu,\tau),\bar{p}(\mu,\tau)),
$$
through the Hamiltonian flow in $\mu$ of $\chi$: 
\begin{equation}\label{eq:sysflotdechi}
\left\{\begin{split}
\frac{\rd\bar{q}(\mu,\tau)}{\rd\mu}&=\left(\partial_{\bar{p}}\chi\right)(\bar{q}(\mu,\tau),\bar{p}(\mu,\tau),\mu,\tau),
\\ \frac{\rd\bar{p}(\mu,\tau)}{\rd\mu}&=-\left(\partial_{\bar{q}}\chi\right)(\bar{q}(\mu,\tau),\bar{p}(\mu,\tau),\mu,\tau),
\end{split}\right.
\end{equation}
with initial conditions $(\bar{q}(0,\tau),\bar{p}(0,\tau))=(\overline{Q},\overline{P}).$ If $\mu$ is small enough, then for all $\tau$ and for $k=1,2$,
\begin{equation}\label{eq:small}
\left(\Phi_{(\mu,\tau)}^\chi\right)^{-1}:\R\times[a/k;\infty[\to\R\times[a/2k;\infty[.
\end{equation}
At the end of the proof we will take $\mu=\varepsilon$, which means $p_*=1/\varepsilon$ has to be large enough.
For all $\mu\leq \varepsilon$ and all $\sigma$ small enough so that $\Phi_\sigma^{\overline H}([a,+\infty[)\subset [a/2, +\infty[$, we define for $(\overline Q, \overline P)\in\R\times [a/2, +\infty[$:
$$
\psi_{(\mu,\sigma)}(\overline Q, \overline P):=\left(\left(\Phi_{(\mu,\tau)}^\chi\right)^{-1}\circ\Phi_\sigma^{\overline{H}}\circ\Phi_{(\mu,0)}^\chi\right)(\overline Q, \overline P).
$$
We also define 
$$
(\overline{Q}(0), \overline{P}(0))=\left(\Phi_{(\mu,0)}^\chi\right)^{-1}(\bar q(0), \bar p(0))\ \mathrm{and}\ 
(\overline{Q}(\sigma), \overline{P}(\sigma))=\psi_{(\mu,\sigma)}(\overline{Q}(0), \overline{P}(0)).
$$ 
Since $\psi_{(\mu,\sigma)}$ is symplectic, there exists a Hamiltonian $\widetilde{H}(\overline{Q},\overline{P},\mu,\tau)$ of which $\psi_{(\mu,\sigma)}$ is the flow with respect to  $\sigma,$ with $\mu$ fixed:
$$\frac{\rd}{\rd\sigma}f\circ\psi_{(\mu,\sigma)}=\left\{f;\widetilde{H}\right\}\circ\psi_{(\mu,\sigma)},\qquad\forall f:\R\times\R^+\to\R.$$
We will write $\psi_{(\mu,\sigma)}=\Phi_{\sigma}^{\widetilde{H}}.$
The expression for $\widetilde{H}$ is known (\cite{cary, lm}):
\begin{equation}\label{eq:defHtilde}
\widetilde{H}=\overline{H}\circ\Phi_{(\mu,\tau)}^\chi-\int_0^\mu\partial_\sigma\chi\circ\Phi_{(\mu',\tau)}^\chi\rd\mu'.
\end{equation}
We obtain the result by constructing $\chi$ as a polynomial of degree $K$ in $\mu,$
$$
\chi(\bar{q},\bar{p},\mu,\tau)=\sum_{n=0}^K\mu^n\chi_{n+1}(\bar{q},\bar{p},\tau),
$$
for each $K\in\N^*,$  in such a way that for $\mu=\varepsilon,$
\begin{equation}\label{eq:theoaverag_pt2}
\frac{\rd}{\rd\sigma}\overline{P}(\sigma)=-\partial_{\overline{Q}}\widetilde{H}\left(\overline{Q}(\sigma),\overline{P}(\sigma),\varepsilon,\tau\right)=\rO(\varepsilon^{K+1});\ \sup_{\sigma\geq 0}|\overline{P}(\sigma)-\bar{p}(\sigma)|=\rO(\varepsilon^2).
\end{equation}
To obtain this, the smooth functions $\chi_n$ will be chosen so that, when $\mu=\varepsilon,$ $\widetilde{H}$ is independent of $\overline{Q}$ up to terms of order $\varepsilon^{K+1}.$
Clearly, for all $f\in C^{\infty}(\R\times\R^+,\R)$ and for all $\tau\geq 0, \bar{q},\overline Q\in \R, \bar{p},\overline P>0$, the functions
$$
\mu\mapsto f\circ\Phi_{(\mu,\tau)}^\chi(\overline{Q},\overline{P})\qquad\textrm{and}\qquad\mu\mapsto f\circ\left(\Phi_{(\mu,\tau)}^\chi\right)^{-1}(\bar{q},\bar{p})
$$
are $\mathcal{C}^{\infty}.$ Indeed, one easily sees that,
with $\Delta_\chi:=\{\cdot;\chi\}+\partial_\mu,$
$$
\partial_\mu^k\left(f\circ\Phi_{(\mu,\tau)}^\chi\right)(\overline{Q},\overline{P})=\left(\Delta_\chi^k f\right)\circ\Phi_{(\mu,\tau)}^\chi(\overline{Q},\overline{P}),\qquad\forall k\in\N,
$$ 
A similar argument works for $\mu\mapsto f\circ\left(\Phi_{(\mu,\tau)}^\chi\right)^{-1}(\bar{q},\bar{p})$.
Hence, such functions can be expanded to any order $K,$ with  smooth coefficients in $(\bar{q},\bar{p},\tau).$ 

We now take $\mu=\varepsilon$ in~\eqref{eq:defHtilde}.  With $\overline{H}$ as in~\eqref{eq:defHbar}, $\widetilde{H}$ is $\mathcal{C}^{\infty}$ in $\varepsilon.$
We introduce the notation
\begin{equation}\label{eq:defHtildeK}
\overline{H}=\overline{H}_0+\varepsilon^2\overline{H}_2,\qquad\widetilde{H}=\sum_{n=0}^K\varepsilon^n\widetilde{H}_n+\rO(\varepsilon^{K+1})=\widetilde{H}^K+\rO(\varepsilon^{K+1}),
\end{equation}
$$
f\circ\left(\Phi_{(\varepsilon,\tau)}^\chi\right)^{-1}=\sum_{n=0}^K\varepsilon^nT_nf+\rO(\varepsilon^{K+1}).
$$
Taking the derivative of~\eqref{eq:defHtilde} with respect to  $\varepsilon$ and composing with $\left(\Phi_{(\varepsilon,\tau)}^\chi\right)^{-1}$ on the right, one finds
$
\partial_\sigma\chi=\partial_\varepsilon\overline{H}+\left\{\overline{H},\chi\right\}-\partial_\varepsilon\widetilde{H}\circ\left(\Phi_{(\varepsilon,\tau)}^\chi\right)^{-1}.
$
Since $\partial_\sigma=\varepsilon\partial_\tau,$ one then easily identifies the terms in $\varepsilon^n$ for $n\in[\![0;K-1]\!],$ on both sides of this equation. For $n=0$, one finds
$
\partial_{\bar{q}}\chi_1=-\widetilde{H}_1/\bar p$, since $\overline{H}_1=0$, $\overline{H}_0={\bar p^2}/2$ (see~\eqref{eq:defHbar}). 
Since we wish to choose $\widetilde{H}_1$ to be independent of $\bar{q},$ averaging this equation over $\bar{q}$ implies $\widetilde{H}_1=0,$ and hence $\partial_{\bar{q}}\chi_1=0;$
one chooses the particular solution $\chi_1=0.$ For $n\in[\![1;K-1]\!],$
one has
\begin{eqnarray*}
\partial_\tau\chi_{n} &=&(n+1)\overline{H}_{n+1}+\left\{\overline{H}_{0};\chi_{n+1}\right\}+\sum_{k=0}^{n-1}\left\{\overline{H}_{n-k};\chi_{k+1}\right\}
\\ & &\qquad\qquad\qquad-(n+1)T_0\widetilde{H}_{n+1}-\sum_{k=1}^{n}\dbinom{n}{k}(n-k+1)T_k\widetilde{H}_{n-k+1},
\end{eqnarray*}
where $\overline{H}_k=0$ when $k\notin\{0;2\}.$ Since $T_0=Id$ one finds,
\begin{eqnarray}
\label{eq:def_chin+1} \partial_{\bar{q}}\chi_{n+1}&=&\frac{1}{\bar{p}}\left[(n+1)\left(\overline{H}_{n+1}-\widetilde{H}_{n+1}\right)-\partial_\tau\chi_{n}+\sum_{k=0}^{n-1}\left\{\overline{H}_{n-k};\chi_{k+1}\right\}\right.\nonumber\\  &&\qquad\qquad\qquad\qquad\left.-\sum_{k=1}^{n}\dbinom{n}{k}(n-k+1)T_k\widetilde{H}_{n-k+1}\right].
\end{eqnarray}
For $k\in[\![1;n]\!],$ the coefficients $T_kf$  depend only on the  $\chi_j$ with  $j\in[\![1;n]\!].$ Indeed, we already know $T_0=Id$ and, in addition, the equality
$$
\frac{\rd}{\rd\mu}\left(f\circ\left(\Phi_{(\mu,\tau)}^\chi\right)^{-1}\right)=-\left\{f\circ\left(\Phi_{(\mu,\tau)}^\chi\right)^{-1};\chi\right\}
$$
implies, after identification of the terms of order $\mu^n$ for  $n\in\N,$
$$
(n+1)T_{n+1}f=-\sum_{k=0}^n\left\{T_{n-k}f;\chi_{k+1}\right\}.
$$ 
After averaging over $\bar q$ in~\eqref{eq:def_chin+1}, one can therefore recursively construct the coefficients $\chi_n$ in such a way that the coefficients $\widetilde{H}_n, n\in[\![0;K]\!],$ don't depend on  $\overline{Q}.$
This yields the first part of~\eqref{eq:theoaverag_pt2}.

To show the second part of \eqref{eq:theoaverag_pt2}, we note that, since $\chi_1=0$, \eqref{eq:sysflotdechi},~\eqref{eq:small}  and the definition of $\overline{P}(\sigma)$ imply
\begin{eqnarray}\label{eq:estimdiffpbar}
\bar{p}(\sigma)-\overline{P}(\sigma) &=&-\int_0^\varepsilon\left(\mu\partial_{\bar{q}}\chi_2(\bar{q}(\mu,\tau),\bar{p}(\mu,\tau),\tau)+\rO(\mu^2)\right)\rd\mu.
\end{eqnarray}
Now,~\eqref{eq:def_chin+1} for $n=1$ reads
$
\partial_{\bar{q}}\chi_{2}=\left[2\left(\overline{H}_{2}-\widetilde{H}_{2}\right)\right]/\bar{p},
$
since $\chi_1=0.$ Integrating this equality with respect to  $\bar{q}$ yields
$0=-\frac{2}{\bar{p}}\widetilde{H}_{2}$
since $\overline{H}_{2}$ has zero average and   $\widetilde{H}_{2}$ is to be chosen independent of $\bar{q}.$ As a result,
$$
\partial_{\bar{q}}\chi_{2}=\frac{2}{\bar{p}}\overline{H}_{2}=\frac{2\lambda}{\bar{p}}V.
$$
The hypotheses on $V$ together with~\eqref{eq:small} therefore imply
$
\sup_{\sigma\geq 0}\left|\bar{p}(\sigma)-\overline{P}(\sigma)\right|=\rO(\varepsilon^2).
$
Similarly,
$
\sup_{\sigma\geq 0}\left|\bar{q}(\sigma)-\overline{Q}(\sigma)\right|=\rO(\varepsilon^2).
$
 One  concludes the proof of the theorem by coming back to the definition of $\bar{p}$ (see \eqref{eq:defphiI}):
$$
p(t)-p(0)=p_*\left(\bar{p}(\sigma)-\bar{p}(0)\right)=\varepsilon^{-1}\left(\bar{p}(\sigma)-\bar{p}(0)\right).
$$
This difference can be decomposed in three terms 
$$
\left|\bar{p}(\sigma)-\bar{p}(0)\right|\leq\left|\bar{p}(\sigma)-\overline{P}(\sigma)\right|+\left|\overline{P}(\sigma)-\overline{P}(0)\right|+\left|\overline{P}(0)-\bar{p}(0)\right|.
$$
Using~\eqref{eq:theoaverag_pt2}, one concludes there exist constants $c_K$ et $C_K$ such that
\begin{eqnarray}\label{eq:pestim}
\left|p(t)-p(0)\right|&\leq& \varepsilon^{-1}c_K\left(\varepsilon^2+\varepsilon^{K+1}\sigma+\varepsilon^2\right)
\leq C_K\varepsilon
\end{eqnarray}
for all $0\leq t=\varepsilon\sigma\leq\sigma_*\varepsilon^{-K+2}.$

\qed

\noindent{\em Proof of Theorem~\ref{co:approxp}.} 
In view of the definitions of $\bar{q},$ $\bar{p}$ and $\sigma$ (see~\eqref{eq:defphiI}), it suffices to show that, for all $0\leq t=\varepsilon\sigma\leq\sigma_*,$ one has 
\begin{equation}\label{eq:goal}
\bar{p}(\sigma)=\bar{p}(0)-\varepsilon^2\frac{\lambda}{\bar{p}(0)}\left(V\left(\bar{q}(0)+\bar{p}(0)\sigma,\tau\right)-V(\bar{q}(0),0)\right)+\rO(\varepsilon^{3}).
\end{equation}
Let $K\in\N,$ $K\geq 3.$ From~\eqref{eq:estimdiffpbar}, using 
$
\overline{P}(\sigma)-\overline{P}(0)=\rO(\varepsilon^{K+1}\sigma),
$
\begin{eqnarray}
\bar{p}(\sigma)-\bar{p}(0)&=&\bar p(\sigma)-\overline{P}(\sigma)+\overline{P}(\sigma)-\overline{P}(0)+\overline{P}(0)-\bar p(0)\nonumber\\
&=&-\int_0^\varepsilon2\lambda\mu\left(\frac{V(\bar{q}(\mu,\tau),\tau)}{\bar{p}(\mu,\tau)}-
\frac{V(\bar{q}(\mu,0),0)}{\bar{p}(\mu, 0)}\right)\rd\mu\nonumber\\
&\ &\qquad\qquad\qquad\qquad\qquad+\rO(\varepsilon^3)+\rO(\varepsilon^{K+1}\sigma).\label{eq:avantderniere}
\end{eqnarray}
Here $\bar q(0,\tau)=\overline Q(\sigma)$ and $\bar p(0,\tau)=\overline P(\sigma)$.
In addition, since $\chi_1=0,$ it follows from~\eqref{eq:small} for all $\sigma$ and all $0\leq \mu\leq \varepsilon$ that
\begin{eqnarray*}
\bar{q}(\sigma)-\overline{Q}(\sigma)&=&\int_0^\varepsilon\partial_{\bar{p}}\chi(\bar{q}(\mu,\tau),\bar{p}(\mu,\tau),\mu,\tau)\rd\mu=\rO(\varepsilon^2),
\\ \bar{q}(\sigma)-\bar{q}(\mu,\tau)&=&\rO(\varepsilon^2),
\qquad\bar{p}(\sigma)-\bar{p}(\mu,\tau)=\rO(\varepsilon^2).
\end{eqnarray*}
Inserting this in~\eqref{eq:avantderniere}, and using $\bar p(\sigma)-\bar p(0)=\rO(\varepsilon^2)$ (see~\eqref{eq:pestim}), one finds
\begin{eqnarray*}
\bar{p}(\sigma)-\bar{p}(0)&=&-\int_0^\varepsilon2\lambda\mu\left(\frac{V(\bar{q}(\sigma),\tau)}{\bar{p}(\sigma)}-\frac{V(\bar{q}(0),0)}{\bar{p}(0)}\right)\rd\mu+\rO(\varepsilon^3)+\rO(\varepsilon^{K+1}\sigma)
\\ &=&-\varepsilon^2\lambda\left(\frac{V(\bar{q}(\sigma),\tau)}{\bar{p}(\sigma)}-\frac{V(\bar{q}(0),0)}{\bar{p}(0)}\right)+\rO(\varepsilon^3)+\rO(\varepsilon^{K+1}\sigma)
\\ &=&-\frac{\varepsilon^2\lambda}{\bar{p}(0)}\left(V(\overline{Q}(\sigma),\tau)-V(\bar q(0),0)\right)+\rO(\varepsilon^3)+\rO(\varepsilon^{K+1}\sigma).
\end{eqnarray*}
Finally, the definition of $\widetilde{H}^K$ (equation~\eqref{eq:defHtildeK}) implies
\begin{eqnarray*}
\overline{Q}(\sigma)&=&\overline{Q}(0)+\int_0^\sigma\partial_{\overline{P}}\widetilde{H}^K(\overline{P}(\sigma'),\varepsilon,\varepsilon\sigma')\rd\sigma'+\rO(\varepsilon^{K+1}\sigma)
\\ &=&\overline{Q}(0)+\int_0^\sigma\partial_{\overline{P}}\widetilde{H}^K(\overline{P}(0),\varepsilon,\varepsilon\sigma')\rd\sigma'+\rO(\varepsilon^{K+1}\sigma)
\\ &=&\bar{q}(0)+\int_0^\sigma\partial_{\overline{P}}\widetilde{H}^K(\bar{p}(0),\varepsilon,\varepsilon\sigma')\rd\sigma'+\rO(\varepsilon^{2})+\rO(\varepsilon^{K+1}\sigma).
\end{eqnarray*}
So
\begin{eqnarray}\label{eq:democoro}
\bar{p}(\sigma)&=&\bar{p}(0)-\frac{\varepsilon^2\lambda}{\bar{p}(0)}\left(V\left(\bar{q}(0)+\int_0^\sigma\partial_{\overline{P}}\widetilde{H}^K(\bar{p}(0),\varepsilon,\varepsilon\sigma')\rd\sigma',\tau\right)-V(\bar{q}(0),0)\right)\nonumber\\
&&\qquad\qquad\qquad+\rO(\varepsilon^{3})+\rO(\varepsilon^{K+1}\sigma).
\end{eqnarray}
To conclude the proof, one fixes $K=3.$ By definition
$$
\widetilde{H}^3=\widetilde{H}_0+\varepsilon\widetilde{H}_1+\varepsilon^2\widetilde{H}_2+\varepsilon^3\widetilde{H}_3,
$$
and we already know  $\widetilde{H}_0=\overline{H}_0=\overline{P}^2/2$ and $\widetilde{H}_1=\widetilde{H}_2=0.$ Considering~\eqref{eq:def_chin+1} with  $n=2$ yields
$
\partial_{\bar{q}}\chi_3=\frac{1}{\bar{p}}\left(-3\widetilde{H}_3-\partial_\tau\chi_2\right).
$
Averaging over $\bar{q}$ implies $\widetilde{H}_3=0,$ since $\chi_2$ can be chosen of zero average. Equation~\eqref{eq:democoro} then yields~\eqref{eq:goal}.
\qed

Equation~\eqref{eq:democoro} implies the following corollary.
\begin{corollary}\label{co:thelastone_yeahbaby!}
Under the hypothesis of Theorem~\ref{th:averaging}, for all $K\in\N,$ $K\geq 3,$ and all $0\leq t\leq\sigma_*p_*^{K-3},$
$$
p(t)=p_{ap}^K(t)+\rO\left(p_*^{-2}\right)
$$
where
$$
p_{ap}^K(t)=p(0)-\frac{\lambda}{p(0)}\left(V\left({q}(0)+p_*\int_0^{t}\partial_{\overline{P}}\widetilde{H}^K\left(\frac{p(0)}{p_*},p_*^{-1},t'\right)\rd t',t\right)-V\left({q}(0),0\right)\right).$$
\end{corollary}


\section{Kicked rotors}\label{s:kicked}
We now turn to the kicked system described in~(\ref{eq:kicked})-(\ref{eq:kickfloquet}) and consider $(q_n, p_n)=\Phi^n(q_0,p_0)\in\T^d\times\R^d$, so that 
\begin{equation}\label{eq:kickiterate}
\left.\begin{split}
p_{n+1}&=p_n-\lambda\nabla v(q_n),
\\ q_{n+1}&=q_n+p_{n+1}.
\end{split}\right\}
\end{equation}
\begin{figure}
\begin{center}
\psfrag{3}{{\tiny $10^3$}}
\psfrag{4}{{\tiny $10^4$}}
\psfrag{5}{{\tiny $10^5$}}
\psfrag{6}{{\tiny $10^6$}}
\psfrag{7}{{\tiny $10^7$}}
\psfrag{8}{{\tiny $10^8$}}
\psfrag{9}{{\tiny $10^9$}}
\psfrag{11}{ }
\psfrag{13}{{\tiny $10^{13}$}}
\psfrag{15}{ }
\psfrag{17}{{\tiny $10^{17}$}}
\psfrag{19}{ }
\psfrag{21}{{\tiny $10^{21}$}}
\psfrag{v8}{{\tiny $\left\|p_0\right\|=1.36$}}
\psfrag{v3}{{\tiny $\left\|p_0\right\|=0.82$}}
\psfrag{v0}{{\tiny $\left\|p_0\right\|=0.50$}}
\psfrag{power1}{{\tiny $\sim t$}}
\psfrag{power2}{{\tiny $\sim t^3$}}
\psfrag{power3}{{\tiny $\sim t$}}
\psfrag{power4}{{\tiny $\sim t^3$}}
\psfrag{t}{{\scriptsize$t$}}
\psfrag{q_t2}{{\scriptsize$\left<\left\|q(t)\right\|^2\right>$}}
\psfrag{p_t2}{{\scriptsize$\left<\left\|p(t)\right\|^2\right>$}}
\psfrag{q_t}{{\scriptsize$\left< q(t)^2\right>$}}
\psfrag{p_t}{{\scriptsize$\left< p(t)^2\right>$}}
\psfrag{1D}{$1d$}
\psfrag{2D}{$2d$}


\includegraphics[height=7.5cm,keepaspectratio=true]{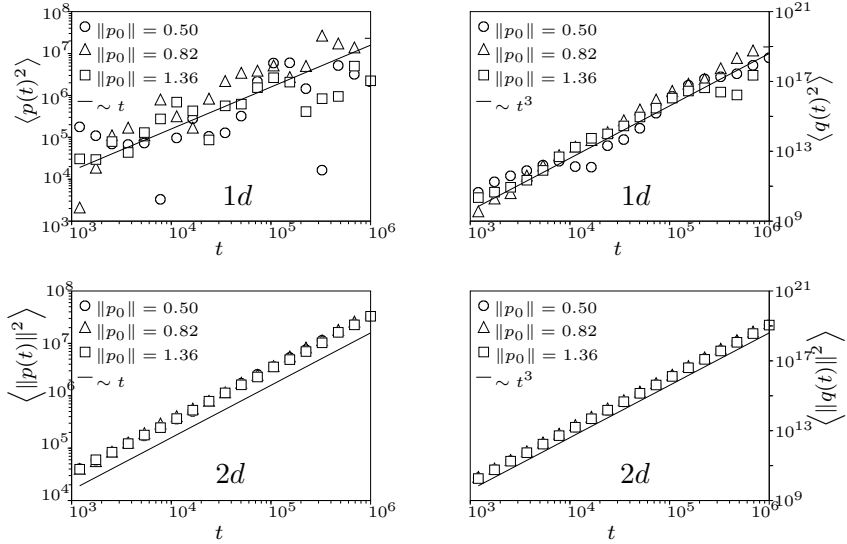}
\end{center}
 \caption{Numerical determined values of $\left<\left\|p(t)\right\|^2\right>$ and $\left<\left\|q(t)\right\|^2\right>$ in one dimension (top) and  two dimensions (bottom) for the model with kicked rotors described by equation~(\ref{eq:kickfloquet}), where $V(q)=\prod_{i=1}^d\cos\left(q_i\right)$ and $\lambda=10.$ The different symbols correspond to different initial conditions, as indicated.}\label{fig:kickedd1}
\end{figure}
Note that here $q_n=q(n)$ and $p_n=p(n)$ since in~(\ref{eq:kicked}) the potential is of period $1$. So in kicked systems, there is exactly one ``scattering event'' per unit time, independent of the particle's speed, whereas in pulsed systems the number of kicks per unit time increases linearly with $\|p\|$ as the particle speeds up. 
So in kicked systems, the momentum change $\Delta_{\mathrm{kick}}\|p\|$ is of order $\lambda$ in a time interval of order $1$, independently of the size of $\|p\|$. On the other hand, in pulsed systems of dimension $d>1$, as a result of (\ref{eq:normprw}), supposing the $\beta_n^{(1)}$ are i.i.d., the typical momentum change $\Delta_{\mathrm{puls}} \|p\|$ of a fast particle with momentum $\|p\|$ in a time interval of order $1$ is
$$
\Delta_{\mathrm{puls}} \|p\|\sim\left<\left(\sum_{n=1}^{\|p\|} \frac{\beta_n^{(1)}}{\|p\|^2}\right)^2\right>^{1/2}\sim\|p\|^{-3/2}.
$$
So
$\Delta_{\mathrm{puls}} \|p\|\ll\Delta_{\mathrm{kick}}\|p\|.$
We will now see that, as a result of this, kicked rotors speed up much faster than pulsed ones. From (\ref{eq:kickiterate}) we find
$$
p_n=p_0-\lambda\sum_{\ell=0}^{n-1} (\nabla V)(q_\ell).
$$
Since in kicked systems $\Delta p_n$ is of order $\lambda$, independently of the size of $\|p\|$, the second equation in (\ref{eq:kickiterate}) suggests that, for large $\lambda$, these successive large momentum changes will lead to a randomization of the position of the system on the torus, allowing one to think of the $q_n$ as uniformly distributed on the torus, and with short temporal correlations. This implies $\left<(p_n-p_0)^2\right>\sim n$ or equivalently
$
\left<\|p(t)\|^2\right>\sim t.
$
In dimension $d=1,$ this immediately implies $\left<\left\|q(t)\right\|^2\right>\sim t^3.$ In dimension $d>1,$ the change of direction $e_{n+1}-e_n,$ with $e_n:=p_n/\left\|p_n\right\|,$ is of order $\left\|p_n\right\|^{-1}\sim n^{-1/2}$, so that 
$\left<\left\|e_{n+m}-e_n\right\|\right>\sim 1$ for $m\sim n$.
Writing, as in section~\ref{s:rw}, $M_k$ for the number of collisions needed by the particle to turn $k$ times over a macroscopic angle, we find $M_k\sim 2^k$ and 
\begin{eqnarray*}
q_{M_{k+1}}&=&q_{M_k}+\left(M_{k+1}-M_k\right)p_{M_k}
\\ &=&q_{M_k}+M_k^{3/2}e_{M_k}.
\end{eqnarray*}
This yields $\left<\|q_{M_k}\|^2\right>\sim M_k^3$ and, interpolating between the $M_k,$ we therefore finally find $\left<\|q(t)\|^2\right>\sim t^3$ even in dimension $d\geq 2$. The reason for this is clear: the particle speeds up considerably faster for kicked systems (where $<\|p(t)\|^2>\sim t$) than for pulsed ones (where $<\|p(t)\|^2>\sim t^{2/5}$) and hence turns much less.
These asymptotics are in agreement with the results of  numerical computations displayed in Figure~\ref{fig:kickedd1}.



\bibliographystyle{alpha}
\bibliography{biblio}

\end{document}